\documentclass[prb,aps,english,onecolumn,notitlepage]{revtex4-1}
\usepackage{epsfig}
\usepackage{amsmath}
\usepackage{mathtools}
\usepackage{sidecap}
\usepackage[figuresleft]{rotating}
\usepackage{caption}
\usepackage{float}
\usepackage{subfigure}
\usepackage{color,pstcol}
\usepackage{graphicx}

\begin{document}
\title{
Characterizing a high spin magnetic impurity via Andreev reflection spectroscopy} 
  \author{Subhajit Pal}
 \author{Colin Benjamin} \email{colin.nano@gmail.com}\affiliation{School of Physical Sciences, National Institute of Science Education \& Research, HBNI, Jatni-752050, India }


\begin{abstract}
 The ground state properties of a high spin magnetic impurity and its interaction with an electronic spin are probed via Andreev reflection. We see that through the charge and spin conductance one can effectively estimate the interaction strength, the ground state spin and magnetic moment of any high spin magnetic impurity. 
 We show how a high spin magnetic impurity at the junction between a normal metal and superconductor can contribute to superconducting spintronics applications. Particularly, while spin conductance is absent below the gap for   Ferromagnet-Insulator-Superconductor junctions we show that in the case of a Normal metal-High spin magnetic impurity-Normal Metal-Insulator-Superconductor (NMNIS) junction it is present. Further, it is seen that pure spin conduction can exist without any accompanying charge conduction in the NMNIS junction. 
 \end{abstract}
\maketitle
\section{Introduction}
Andreev reflection, the process of an incident electron (hole) being retro-reflected as a hole (electron) at an interface with a superconductor has been used as a spectroscopic probe for the amount of polarization in ferromagnet's\cite{soulen,been}, to determine the pairing symmetry of High T$_{c}$\cite{tsui,colin} and Ferromagnetic superconductors\cite{colin} among many other uses. 
High spin magnetic impurities (HSM)  have been {realized} around for 20 years or more. The main purpose of research in these is to design molecular magnets however another important direction is towards controlled concoction of stable high spin molecular complexes and their complete characterization, as regards the ground state spin properties of these magnetic impurities. For example, how these magnetic impurities interact with electron spin, their ground state spin and magnetic moment have been only deciphered via spectroscopic probes like- pulsed electron paramagnetic resonance (EPR)\cite{schweiger}, electron nuclear double resonance spectroscopy (ENDOR)\cite{kurreck} or electron spin echo envelope modulation (ESEEM) techniques. The problem with the aforesaid spectroscopic tools like EPR, ENDOR, EESEM, etc are that they only probe one of the features like EPR and ENDOR can only probe the ground state spin while magnetic moment can be done only with EESEM techniques. In Ref.~\cite{Bur} different electron transport methods to probe the magnetic anisotropy of HSM has been discussed. Gate spectroscopy is one of them, which quantifies the longitudinal magnetic anisotropy of the HSM in different redox states. However in our work, we probe spin, magnetic moment and exchange interaction of HSM, while Ref.~\cite{Bur} probes axial and transverse anisotropy which are not considered in our work. So, our work is complementary to Ref.~\cite{Bur}, looking at both one can get the complete picture of HSM.  New techniques like infrared spectroscopy and magnetic circular dichroism spectroscopy have also been invoked recently\cite{ncomm} to fully characterize HSM's, but as already mentioned the multiplicity of spectroscopic  techniques increases the complexity of the problem. We in this work make an attempt to show that Andreev reflection spectroscopy can be not only an alternative but also a genuine technique to characterize the spin, magnetic moment and the exchange interaction of a HSM thus reducing the complexity involved. 

In this work we not only show that Andreev reflection\cite{BTK} can be an excellent tool to probe these aforesaid aims, but also explore the possibilities of utilizing high spin magnetic impurities in superconducting spintronics applications. High spin magnetic impurities(HSM) are of great importance in molecular spintronics. The Hamiltonian\cite{AJP,mauri} used to describe a HSM is given by-
\begin{equation}
H_{HSM}=-J_{0}\vec{s}.\vec{S}  
\end{equation}
{where $J_{0}$ is the strength of the exchange interaction between the electron with spin $\vec{s}$ and a magnetic impurity with spin $\vec{S}$. Explicitly in terms of spin raising and lowering operators for electron as well as magnetic impurity we can write-
\begin{equation}
\vec s.\vec S=s^{z}S^{z}+\frac{1}{2}(s^{-}S^{+}+s^{+}S^{-})\nonumber 
\end{equation}
where $s^{\pm} = s_{x}\pm is_{y}$ are the spin raising and lowering operators for electron and $S^{\pm} = S_{x}\pm iS_{y}$ are the spin raising and lowering operators for HSM.}
 The above model for a magnetic impurity in a Andreev setting matches quite well with solid-state scenarios such as
seen in  1D quantum wires or graphene  with an embedded magnetic impurity or quantum dot\cite{palma}. 
Electrons interact with HSM via   $-J_{0}\vec{s}.\vec{S}$, where $J_{0}$ being the strength of the exchange interaction, $\vec{s}$ is the electronic spin and $\vec{S}$ is the spin of the magnetic impurity. $J_{0}(=\frac{\hbar^2 k_{F}J}{m^{\star}})$, with $J$ being the relative magnitude of the exchange interaction which ranges from $0-5$ in this work, $m^{\star}$ is the electronic mass and Fermi wavevector $k_{F}$ is obtained from the Fermi energy $E_{F}$ which is the largest energy scale in our system $1000\Delta$, $\Delta$ being the superconducting gap for a widely used s-wave superconductor like Aluminium is $0.17$ meV. Substituting this value of the Fermi wavevector so obtained in the formula for $J_{0}$ we get $J_{0}=  0.160$ eV (if $J=1$). The exact setting we will use is shown in Fig.~1, it shows  a  HSM at $x=0$ and a delta potential barrier at $x=a$. In the regions I ($x<0$) and II ($0<x<a$) there are two  normal metals while for $x>a$ there is a s-wave superconductor. We study the Andreev reflection enabled transport across this junction especially concentrating on  below the gap regime of the superconductor. We consider unpolarized  electrons incident at the junction and show how the differential charge and spin conductance vary with the impurity spin S and z-component of the impurity spin $m'$ respectively. From the values of the differential spin and charge conductance we can get the exact values of the exchange interaction strength, ground state spin and magnetic moment of the magnetic impurity. We also study the spin  and charge conductance through the junction and its possible applications in superconducting spintronics\cite{eschrig}. Particularly, we focus on conditions for obtaining pure spin conductance in absence of any charge conductance. Thus the twin objectives of this study are: A. To characterize the ground state spin, magnetic moment and exchange interaction of a HSM and B. To exploit quantum transport across this Normal Metal-HSM-Normal Metal-Insulator-Superconductor (NMNIS) junction to design a pure spin conducting device.\\ 
 The topic of magnetic impurity in vicinity of superconductor has been explored before with d-wave superconductor. A point like impurity in a d-wave superconductor forms midgap\cite{mid} states within the superconducting gap. A magnetically doped superconductor has been studied experimentally with a low-temperature scanning tunneling microscope\cite{yazdani}. In Ref.~\cite{Frank} the authors show how magnetic and superconducting interactions can coexist and influence the ground state of a magnetic impurity. When a magnetic impurity is absorbed on the surface of a superconductor, its spin can interact with itinerant electrons (spin $s=1/2$) and with Cooper pair ($s=0$). Normal state electrons tend to screen the magnetic impurity and form a many particle ground state with total spin $S=0$. This effect is called Kondo effect\cite{Frank}. In our case magnetic impurity lies at the interface between metal and superconductor and we neglect e-e interactions. So, we are away from the Kondo regime.     
 
The rest of the paper is organized as follows: in the succeeding section {on Theory}, we first outline a brief sketch of our set-up and provide a theoretical background to our study {with Hamiltonian, wavefunctions and boundary conditions to calculate the different reflection and transmission probabilities. We study the effect of the spin of HSM on these probabilities both for transparent as well as tunneling regimes.} 
The section after Theory deals with the first objective of this work, i.e. characterization, in three separate subsections as well as two extensive tables and three elaborate figures we explain the way to characterize the HSM via only the differential charge and spin conductance. In the next section following 
we study the effect of finite temperature on Andreev transport through HSM. The section after {the effect of finite temperature} deals with the second aim of this work namely applications in superconducting spintronics, we show how we can utilize our set-up to have pure spin conductance in absence of any charge conductance. Finally we conclude our study with a section on conclusion and a perspective on future endeavors. We also provide an Appendix wherein details of the calculations are provided.
\begin{figure}[h]
\centering{\includegraphics[width=.8\linewidth]{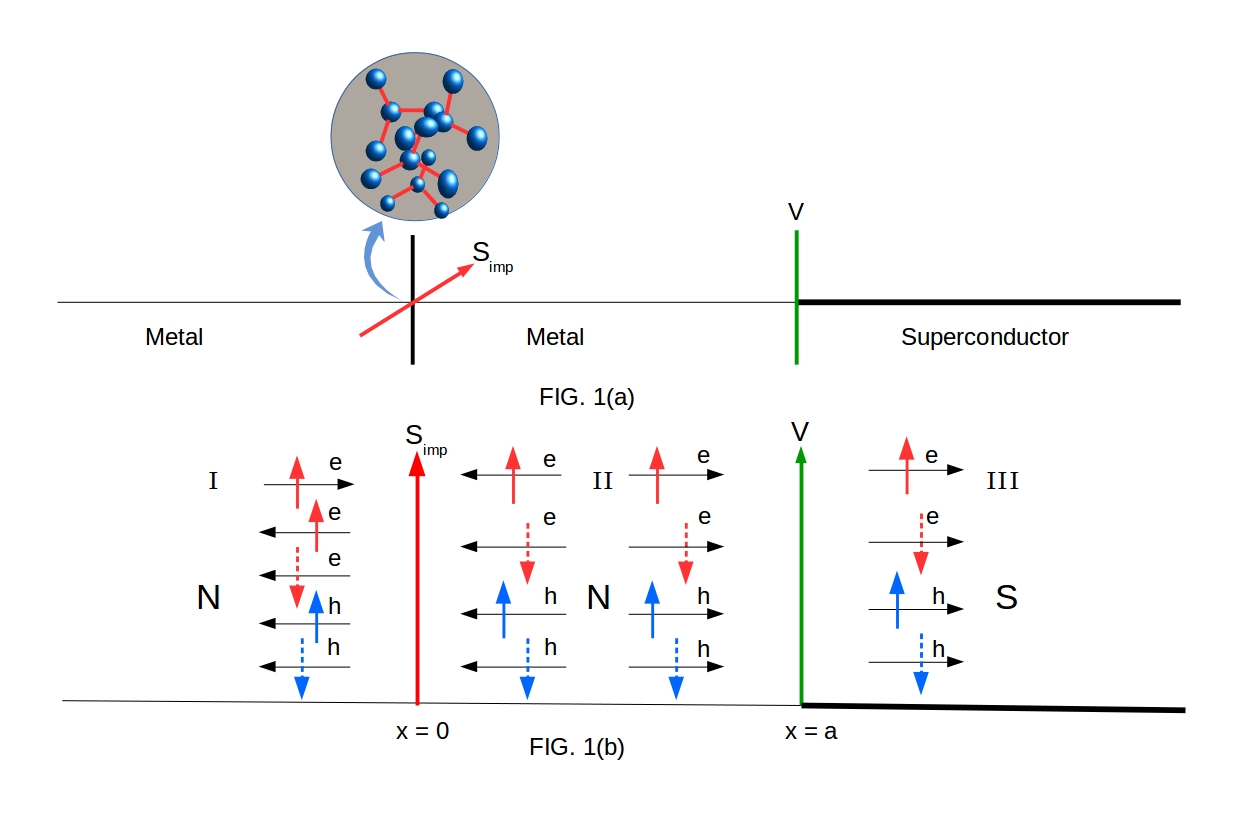}}
\caption{(a) A high spin magnetic impurity with spin S and magnetic moment $m'$ at $x=0$ in a Normal Metal-HSM-Normal Metal-Insulator-Superconductor (NMNIS) junction, (b) The scattering of an up-spin electron incident is shown. Andreev reflection and quasi particle transmission into superconductor are depicted. For details see section II B.}
\end{figure}

\section{Theory}
\subsection{Hamiltonian}
We consider a metal (N)-metal (N)-superconductor (S) junction where there is a HSM between two metallic regions at $(x=0)$ and a $\delta$-like potential barrier exists at metal superconductor interface at $(x=a)$. When an electron with energy E and spin ($\uparrow/\downarrow$) is incident from the normal metal, at the $x=0$ interface it interacts with the HSM through an exchange potential which may induce a mutual spin flip. The electron can be reflected back to region I, or transmitted to region II, with spin up or down. When this transmitted electron is incident at $x=a$ interface it could be reflected back from the interface and there is also possibility of Andreev reflection, i.e., a hole with spin up or down is reflected back to region II. Electron-like and hole-like quasi-particles with spin up or down are transmitted into the superconductor for energies above the gap.

The model Hamiltonian in Bogoliubov-de Gennes formalism of our Normal Metal-Magnetic impurity-Normal Metal-Insulator-Superconductor system is given below:
\begin{eqnarray}
  \begin{bmatrix}
    H\hat{I} & i\Delta \theta(x-a)\hat{\sigma}_{y}  \\
    -i\Delta^{*}\theta(x-a)\hat{\sigma}_{y}   &  -H\hat{I}
  \end{bmatrix} \psi(x)& =& E \psi(x), \mbox{ with } H = p^2/2m^\star + V\delta(x-a) - J_{0}\delta(x)\vec s.\vec S -E_{F},\nonumber\\
&&\mbox{ $\psi$ is a four-component spinor, $\Delta$ is the gap in s-wave superconductor and }\nonumber\\
&&\theta \mbox{   is the Heaviside step function.}
\end{eqnarray}
Further, in $H$ the first term is the kinetic energy of an electron with effective mass $m^\star$, for second term  $V$ is the strength of the $\delta$-like potential at the interface between normal metal and superconductor, the third term describes the exchange interaction of strength $J_{0}$ between the electron with spin $\vec s$ and a magnetic impurity with spin $\vec S$, $\hat{\sigma}$ is the Pauli spin matrix and $\hat{I}$ is unit matrix, $E_F$ being the Fermi energy. We will later use the dimensionless parameter $J=\frac{m^{\star}J_{0}}{\hbar^2 k_{F}}$ as a measure of strength of exchange interaction\cite{AJP} and $Z=\frac{m^{\star}V}{\hbar^2 k_{F}}$ as a measure of interface transparency\cite{BTK}. In our work $Z$ is a dimensionless quantity, while $V$ has the dimension of energy. $Z$ denotes the transparency of the junction, $Z=0$ means completely transparent junction, while $Z>>1$ implies a tunneling junction\cite{b,BTK}. 
\subsection{Wavefunctions}
The wave functions of the different region of the system as shown in FIG. 1(a) and FIG. 1(b) can be written in spinorial form\cite{LINDER}:
\[\psi_{N}^{I}(x)=\begin{bmatrix}
                    1\\
                    0\\
                    0\\
                    0
                  \end{bmatrix}e^{ik_{e}x}\phi_{m'}^{S}+r_{ee}^{\uparrow\uparrow}\begin{bmatrix}
                  1\\
                  0\\
                  0\\
                  0
                 \end{bmatrix}e^{-ik_{e}x}\phi_{m'}^{S}+r_{ee}^{\uparrow\downarrow}\begin{bmatrix}
                 0\\
                 1\\
                 0\\
                 0
                \end{bmatrix}e^{-ik_{e}x}\phi_{m'+1}^{S}+r_{eh}^{\uparrow\uparrow}\begin{bmatrix}
                0\\
                0\\
                1\\
                0
               \end{bmatrix}e^{ik_{h}x}\phi_{m'+1}^{S}+r_{eh}^{\uparrow\downarrow}\begin{bmatrix}
               0\\
               0\\
               0\\
               1
              \end{bmatrix}e^{ik_{h}x}\phi_{m'}^{S},\mbox{for $x<0$}\]
\begin{eqnarray}
\Psi_{N}^{II}(x)=t_{ee}^{'\uparrow\uparrow}\begin{bmatrix}
                                     1\\
                                     0\\
                                     0\\
                                     0
                                     \end{bmatrix}e^{ik_{e}x}\phi_{m'}^{S}+t_{ee}^{'\uparrow\downarrow}\begin{bmatrix}
                                     0\\
                                     1\\
                                     0\\
                                     0
                                    \end{bmatrix}e^{ik_{e}x}\phi_{m'+1}^{S}+b_{ee}^{\uparrow\uparrow}\begin{bmatrix}
                                    1\\
                                    0\\
                                    0\\
                                    0
                                   \end{bmatrix}e^{-ik_{e}(x-a)}\phi_{m'}^{S}+b_{ee}^{\uparrow\downarrow}\begin{bmatrix}
                                   0\\
                                   1\\
                                   0\\
                                   0
                                  \end{bmatrix}e^{-ik_{e}(x-a)}\phi_{m'+1}^{S}\nonumber\\+c_{eh}^{\uparrow\uparrow}\begin{bmatrix}
                                  0\\
                                  0\\
                                  1\\
                                  0
                                 \end{bmatrix}e^{ik_{h}(x-a)}\phi_{m'+1}^{S}+c_{eh}^{\uparrow\downarrow}\begin{bmatrix}
                                 0\\
                                 0\\
                                 0\\
                                 1
                                \end{bmatrix}e^{ik_{h}(x-a)}\phi_{m'}^{S}+a_{eh}^{\uparrow\uparrow}\begin{bmatrix}
                                0\\
                                0\\
                                1\\
                                0
                               \end{bmatrix}e^{-ik_{h}x}\phi_{m'+1}^{S}+a_{eh}^{\uparrow\downarrow}\begin{bmatrix}
                               0\\
                               0\\
                               0\\
                               1
                              \end{bmatrix}e^{-ik_{h}x}\phi_{m'}^{S},\mbox{for $0<x<a$}\nonumber
                              \end{eqnarray}
\[\psi_{S}(x)=t_{ee}^{\uparrow\uparrow}\begin{bmatrix}
                              u\\
                              0\\
                              0\\
                              v
                             \end{bmatrix}e^{iq_{+}x}\phi_{m'}^{S}+t_{ee}^{\uparrow\downarrow}\begin{bmatrix}
                             0\\
                             u\\
                             -v\\
                             0
                             \end{bmatrix}e^{iq_{+}x}\phi_{m'+1}^{S}+t_{eh}^{\uparrow\uparrow}\begin{bmatrix}
                             0\\
                             -v\\
                             u\\
                             0
                             \end{bmatrix}e^{-iq_{-}x}\phi_{m'+1}^{S}+t_{eh}^{\uparrow\downarrow}\begin{bmatrix}
                             v\\
                             0\\
                             0\\
                             u
                             \end{bmatrix}e^{-iq_{-}x}\phi_{m'}^{S},\mbox{for $x>a$}\] \\
 $r_{ee}^{\uparrow\uparrow}$($r_{ee}^{\uparrow\downarrow}$) and $r_{eh}^{\uparrow\uparrow}$($r_{eh}^{\uparrow\downarrow}$) are the corresponding  amplitudes for normal reflection and Andreev reflection with spin up(down). $t_{ee}^{\uparrow\uparrow}$($t_{ee}^{\uparrow\downarrow}$) and $t_{eh}^{\uparrow\uparrow}$($t_{eh}^{\uparrow\downarrow}$) are the corresponding amplitudes for transmission of electron-like quasi-particles and hole-like quasi-particles with spin up(down). $\phi_{m'}^{S}$ is the eigenfunction of magnetic impurity: with its $S^{z}$ operator acting as-
$S^{z}\phi_{m'}^{S} = m'\phi_{m'}^{S}$, with $m'$ being the spin magnetic moment of the HSM.  
For $E>\Delta$(for energies above the gap),the coherence factors are $u^2=\frac{1}{2}\Big[\frac{E+(E^2-\Delta^2)^\frac{1}{2}}{E}\Big]$, $v^2=\frac{1}{2}\Big[\frac{E-(E^2-\Delta^2)^\frac{1}{2}}{E}\Big]$, while the wave-vector in metal is $k_{e,h}=\sqrt{2m^{\star}(E_{F}\pm E)}$ and in superconductor is $q_{\pm}=\sqrt{2m^{\star}(E_{F}\pm \sqrt{E^2-\Delta^2})}$ and for $E<\Delta$(for energies below the gap) the coherence factors are {$u^2=\frac{1}{2}\Big[\frac{E+i(\Delta^2-E^2)^\frac{1}{2}}{\Delta}\Big]$, $v^2=\frac{1}{2}\Big[\frac{E-i(\Delta^2-E^2)^\frac{1}{2}}{\Delta}\Big]$}, while the wave-vector in metal is 
$k_{e,h}=\sqrt{2m^{\star}(E_{F}\pm E)}$ and in superconductor is $q_{\pm}=\sqrt{2m^{\star}(E_{F}\pm i\sqrt{\Delta^2-E^2})}$\cite{BTK}, wherein $E_{F}$ is the Fermi energy, $m^{*}$ is the effective mass of electron in metal and $E$ is the excitation energy of electron above $E_{F}$. In Andreev approximation, which we will use throughout this work, $E_{F}\gg \Delta, E$ we take $k_{e}=k_{h}=q_{+}=q_{-}=k_{F}$.
We impose the boundary conditions on the above wave-functions and solve the resulting $16$ equations and get the different scattering amplitudes: $r_{ee}^{\uparrow\uparrow},r_{ee}^{\uparrow\downarrow},r_{eh}^{\uparrow\uparrow},r_{eh}^{\uparrow\downarrow},t_{ee}^{\uparrow\uparrow},t_{ee}^{\uparrow\downarrow},t_{eh}^{\uparrow\uparrow},t_{eh}^{\uparrow\downarrow}$, see supplementary material. The reflection and transmission probabilities we get are thus-  $R_{ee}^{\uparrow\uparrow}= |r_{ee}^{\uparrow\uparrow}|^{2},R_{ee}^{\uparrow\downarrow}=|r_{ee}^{\uparrow\downarrow}|^{2},R_{eh}^{\uparrow\uparrow}=|r_{eh}^{\uparrow\uparrow}|^{2},R_{eh}^{\uparrow\downarrow}=|r_{eh}^{\uparrow\downarrow}|^{2}$, {$T_{ee}^{\uparrow\uparrow}=(u^2-v^2)|t_{ee}^{\uparrow\uparrow}|^{2}, T_{ee}^{\uparrow\downarrow}=(u^2-v^2)|t_{ee}^{\uparrow\downarrow}|^{2},T_{eh}^{\uparrow\uparrow}=(u^2-v^2)|t_{eh}^{\uparrow\uparrow}|^{2},T_{eh}^{\uparrow\downarrow}=(u^2-v^2)|t_{eh}^{\uparrow\downarrow}|^{2}$}, these are plotted in Figures below.
\begin{figure}[h]  
\includegraphics[width=.88\textwidth]{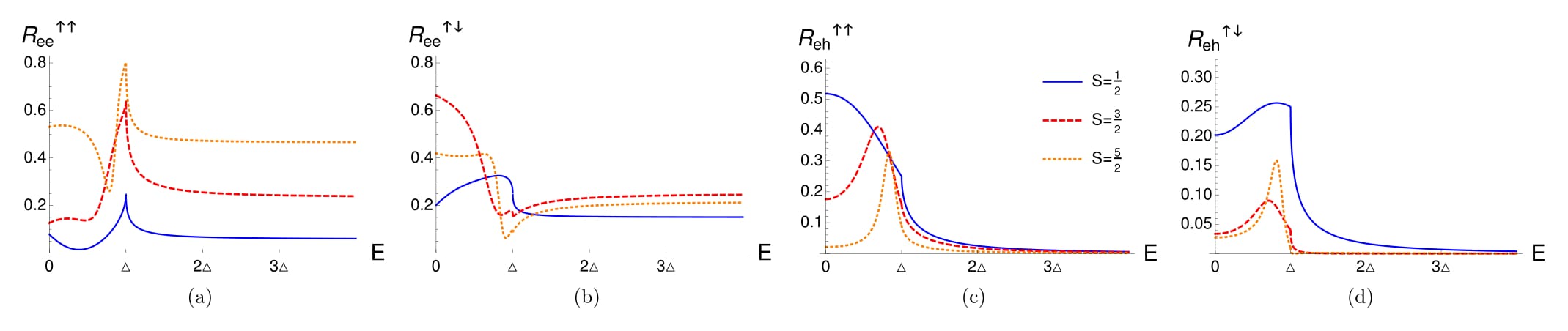}
\caption{a) Normal reflection  probability without flip, b) Normal reflection probability with flip, c) Andreev reflection probability with flip, d) Andreev reflection probability without flip, in the transparent regime. Parameters for all are: \small \sl $J=1.0,Z=0.0,m'=-1/2$}
\end{figure}
\begin{figure}[h]  
\includegraphics[width=.88\textwidth]{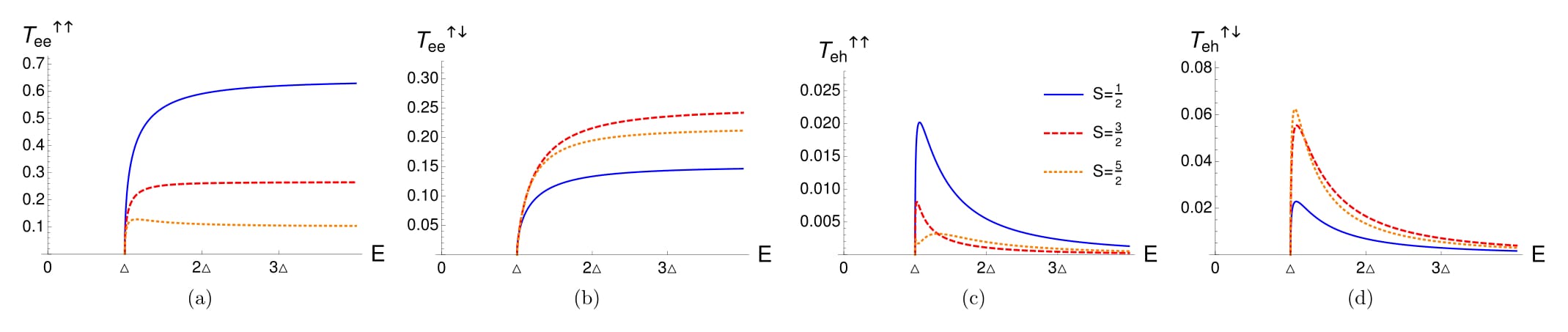}
\caption{a) Electron-like quasi-particle transmission without flip, b) Electron-like quasi-particle transmission with flip, c) Hole-like quasi-particle transmission with flip, d) Hole-like quasi-particle transmission without flip, in the transparent regime. Parameters: \small \sl $J=1.0,Z=0.0,m'=-1/2$}
\end{figure}
\newpage
\begin{figure}[h]  
\includegraphics[width=.88\textwidth]{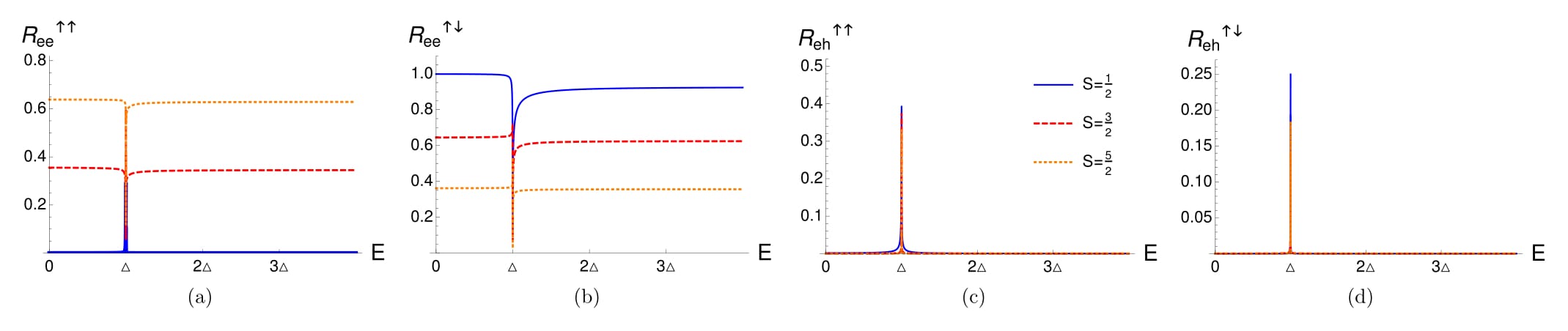}
\caption{a) Normal reflection  probability without flip, b) Normal reflection probability with flip, c) Andreev reflection probability with flip, d) Andreev reflection probability without flip, in the tunneling regime. Parameters for all are: $J=1.0,Z=3.0,m'=-1/2$}
\end{figure}
\begin{figure}[h]  
\includegraphics[width=.88\textwidth]{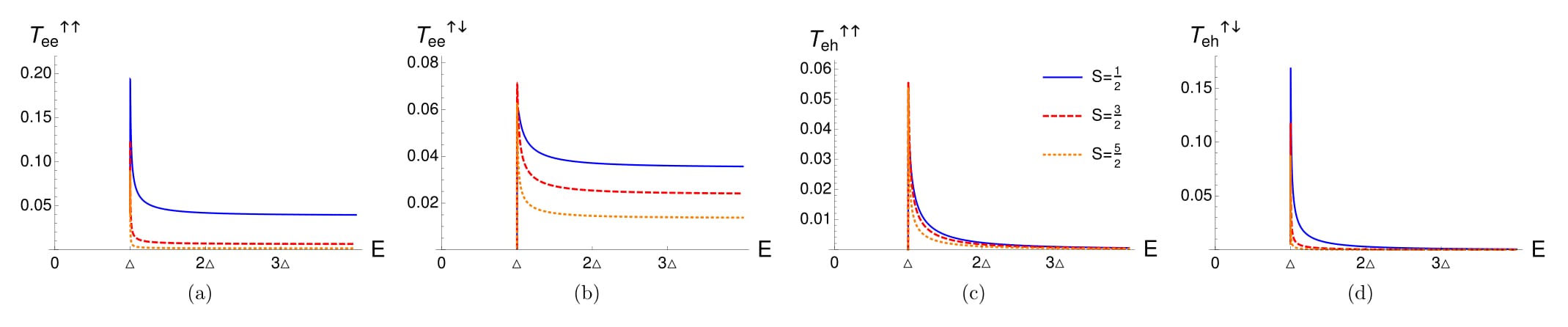}
\caption{a)Electron-like quasi-particle transmission without flip, b) Electron-like quasi-particle transmission with flip, c) Hole-like quasi-particle transmission with flip, d) Hole-like quasi-particle transmission without flip, in the tunneling regime. Parameters: $J=1.0,Z=3.0,m'=-1/2$}
\end{figure}
\subsection{Andreev Reflection in presence of a HSM}
In Fig.~2 we plot the normal and Andreev reflection probabilities with spin-flip or no-flip for different values of the spin of  magnetic impurity $S(1/2,3/2,5/2)$, we fix the magnetic moment of the magnetic impurity- $m'=-1/2$ and we take $Z=0$- the transparent regime. In Fig.~2(a) we see normal reflection probability without spin flip increases with increase of spin of magnetic impurity (S), while in 2(b) normal reflection probability with spin flip shows a mixed behavior it first increases then decreases with increase of S, this is for the entire range of electron excitation energy. Next in Fig.~2(c) we plot Andreev reflection with spin flip, for increasing spin it continuously decreases for both above as well as below the gap. Finally, in Fig.~2(d) we plot the Andreev reflection probability without spin flip, we find both above as well as below the gap, probability decreases as  spin increases. Thus, for a high spin magnetic impurity, normal reflection probability is large but Andreev reflection probability is small. Andreev reflection, hence is inhibited by the spin of magnetic impurity. This behavior, especially for large S, has similarities as well as differences with  seen for a Ferromagnet-insulator-Superconductor junction\cite{been}. So, although it is one of the main aims of high spin magnetic impurity research to design single magnetic impurity magnets, one of the conclusions of our work is that high spin magnetic impurities are not Ferromagnets.

 In Fig.~3, we plot the quasi-particle transmission probabilities for spin flip and without spin flip for same parameter values as in Fig.~2. Since there is no quasi-particle transmission below the gap we will only focus on above the gap regime. In Fig.~3(a) we show that with increase in spin of magnetic impurity (S) the electron-like quasi-particle transmission without flip decreases on the other hand in Fig.~3(b) with increase in spin of magnetic impurity (S) the electron-like quasi-particle transmission with flip increases. Thus high spin magnetic impurities inhibit no flip transmission but boost transmission with spin-flip for electron-like quasi-particle. In Fig.~3(c) and (d) we plot the probability for hole-like quasi-particle transmission, in (c) we see that with increase in spin of magnetic impurity (S) the probability for spin-flip transmission decreases while in (d) we see the opposite. Thus high spin magnetic impurities show opposite behavior for holes, they inhibit spin-flip transmission while giving a boost to no-flip transmission.

 While Figs.~2 and 3 dealt with reflection and transmission in the transparent (Z=0) regime, in Figs.~4 and 5 we deal with the tunneling (Z=3) regime, other parameters remain same. In Fig.~4(a) we see the normal reflection probability in absence of spin flip increases with increase of the spin (S) of magnetic impurity, while the normal reflection probability with spin flip decreases with spin (S). Since in the tunneling regime Andreev reflection by default is inhibited the addition of magnetic impurity enhances normal reflection and further inhibits  Andreev reflection for both spin-flip and no-flip cases. Finally, in Fig.~5 we plot the quasi-particle transmission probabilities. We see in all cases increasing spin (S) of magnetic impurity leads to a continuous decrease of both electron-like and hole-like transmission probabilities. Further, in the tunneling regime the differences between spin states of the HSM are almost obliterated.

Similar to above, if we consider an electron with spin down incident from normal metal region I, and get the different reflection probabilities as follows: $R_{ee}^{\downarrow\uparrow},R_{ee}^{\downarrow\downarrow},R_{eh}^{\downarrow\uparrow},R_{eh}^{\downarrow\downarrow}$, and for quasiparticle transmission above the gap $T_{ee}^{\downarrow\uparrow}, T_{ee}^{\downarrow\downarrow}, T_{eh}^{\downarrow\uparrow}, T_{eh}^{\downarrow\downarrow}$.
As we solve for the scattering amplitudes when electron with spin up is incident, we can also do the same for a spin down incident electron. We do not repeat them here but in analogy to spin up case the wavefunctions can be easily written.
\subsection{Differential charge and spin conductance}
To calculate the  total differential spin conductance, we follow the well established definitions as in Refs.~{\cite{jin,kashiwaya}}. The spin conductance of the set-up as envisaged in Fig.~1 is given by-
\begin{equation}
G_{s}^{0}=G_{s}^{\uparrow}-G_{s}^{\downarrow}\mbox{,  where $G_{s}^{\uparrow}=1+R_{eh}^{\uparrow\uparrow}-R_{eh}^{\uparrow\downarrow}-R_{ee}^{\uparrow\uparrow}+R_{ee}^{\uparrow\downarrow}$ and $G_{s}^{\downarrow}=1+R_{eh}^{\downarrow\downarrow}-R_{eh}^{\downarrow\uparrow}-R_{ee}^{\downarrow\downarrow}+R_{ee}^{\downarrow\uparrow}$}\nonumber
\end{equation}
So,
\begin{equation}
G_{s}^{0}=R_{eh}^{\uparrow\uparrow}-R_{eh}^{\uparrow\downarrow}-R_{ee}^{\uparrow\uparrow}+R_{ee}^{\uparrow\downarrow}-R_{eh}^{\downarrow\downarrow}+R_{eh}^{\downarrow\uparrow}+R_{ee}^{\downarrow\downarrow}-R_{ee}^{\downarrow\uparrow} 
\end{equation}
while the net differential charge conductance is defined as \cite{jin,kashiwaya}-
\begin{equation}
G_{c}^{0}=G_{c}^{\uparrow}+G_{c}^{\downarrow}\mbox{, where $G_{c}^{\uparrow}=1+R_{eh}^{\uparrow\uparrow}+R_{eh}^{\uparrow\downarrow}-R_{ee}^{\uparrow\uparrow}-R_{ee}^{\uparrow\downarrow}$ and $G_{c}^{\downarrow}=1+R_{eh}^{\downarrow\uparrow}+R_{eh}^{\downarrow\downarrow}-R_{ee}^{\downarrow\uparrow}-R_{ee}^{\downarrow\downarrow}$}\nonumber 
\end{equation}
So,
\begin{equation}
G_{c}^{0}=2+R_{eh}^{\uparrow\uparrow}+R_{eh}^{\uparrow\downarrow}-R_{ee}^{\uparrow\uparrow}-R_{ee}^{\uparrow\downarrow}+R_{eh}^{\downarrow\downarrow}+R_{eh}^{\downarrow\uparrow}-R_{ee}^{\downarrow\downarrow}-R_{ee}^{\downarrow\uparrow}
\end{equation}
where $R_{eh}^{\uparrow\uparrow}$ is the probability of Andreev reflection of an electron (spin up)as hole (spin up),
$R_{eh}^{\uparrow\downarrow}$ is the probability of Andreev reflection of an electron (spin up)as hole (spin down),
$R_{ee}^{\uparrow\uparrow}$ is the probability of normal reflection of an electron (spin up)as electron (spin up),
$R_{ee}^{\uparrow\downarrow}$ is the probability of normal reflection of an electron (spin up)as electron (spin down),
$R_{eh}^{\downarrow\downarrow}$ is the probability of Andreev reflection of an electron (spin down)as hole (spin down),
$R_{eh}^{\downarrow\uparrow}$ is the probability of Andreev reflection of an electron (spin down)as hole (spin up),
$R_{ee}^{\downarrow\downarrow}$ is the probability of normal reflection of an electron (spin down)as electron (spin down),
$R_{ee}^{\downarrow\uparrow}$ is the probability of normal reflection of an electron (spin down)as electron (spin up).\\

Now for below the gap ($-\Delta <E< \Delta$) regime and with spin up electron incident, the conservation of probability gives-
$R_{eh}^{\uparrow\uparrow}+R_{eh}^{\uparrow\downarrow}+R_{ee}^{\uparrow\uparrow}+R_{ee}^{\uparrow\downarrow}=1$
and for spin down electron incident, the conservation of probability gives-
$R_{eh}^{\downarrow\uparrow}+R_{eh}^{\downarrow\downarrow}+R_{ee}^{\downarrow\uparrow}+R_{ee}^{\downarrow\downarrow}=1$.\\
Putting these two conditions in the differential conductances we get, for below the gap ($-\Delta <E< \Delta$):
\begin{equation}
G_{s}=\frac{1}{2\pi}\int_{0}^{2\pi} 2(R_{eh}^{\uparrow\uparrow}+R_{ee}^{\uparrow\downarrow}-R_{eh}^{\downarrow\downarrow}-R_{ee}^{\downarrow\uparrow}) d(k_{F}a)= \frac{1}{2\pi}\int_{0}^{2\pi} G_{s}^{0} d(k_{F}a)
\end{equation}
and 
\begin{equation}
G_{c}=\frac{1}{2\pi}\int_{0}^{2\pi} 2(R_{eh}^{\uparrow\uparrow}+R_{eh}^{\uparrow\downarrow}+R_{eh}^{\downarrow\downarrow}+R_{eh}^{\downarrow\uparrow}) d(k_{F}a)= \frac{1}{2\pi}\int_{0}^{2\pi} G_{c}^{0} d(k_{F}a)
\end{equation}

 The differential charge conductance at zero bias ($E=0$) for transparent junction ($Z=0$):
\begin{equation}
 G_{c}^{0}\neq 0
\end{equation}

Similarly, the differential spin conductance at zero bias ($E=0$ or without any voltage bias applied to normal metal region I) for transparent junction ($Z=0$):
\begin{equation}
 G_{s}^{0}\neq 0
\end{equation}

The differential charge conductance in the zero bias $E=0$ limit and in the tunneling $Z\rightarrow Large $ limit,  vanishes:
\begin{equation}
G_{c}^{0}=0
\end{equation}

while the differential spin conductance in the zero bias $E=0$ limit and in the tunneling $Z\rightarrow Large $ limit, is given by:
\begin{equation}
 G_{s}^{0}\neq 0
\end{equation}

Thus, in the tunneling regime $Z\rightarrow Large $ and in absence of any voltage bias $E=0$, we have pure spin conductance for our NMNIS junction in absence of any charge conductance. This is one of the main results of this work. \\

\section{Characterizing the high spin magnetic impurity spin, magnetic moment and interaction strength}
In Table 1 (see Appendix) we tabulate the values of the differential spin and charge conductance for different values of spin ($S$) and magnetic moment ($m'$) of the magnetic impurity in the zero bias limit, this limit implies the differential conductance is nothing but the total conductance. We further address the case for transparent contact ($Z=0$) at the metal-superconductor interface. We see the differential charge conductance ($G_c$) for a particular $S$, varies as a function of the magnitude of magnetic moment but is independent of the sign of magnetic moment. Further, $G_{c}$ at $m'=G_{c}$ at $-m'$. The differential spin conductance on the other hand follows the relation $G_{s}$ at $m'=-G_{s}$ at $-m'$ and can also be negative\cite{tanaka}, further the magnitude of $G_{s}$ increases with increasing magnitude of $m'$ for a particular $S$.
In Table 2 (see Appendix), we tabulate the charge and spin conductance again but for the energy at gap edge limit $E=\Delta$, here we also see that $G_{s}$ at $m'=-G_{s}$ at $-m'$ and the magnitude of $G_{s}$ increases with increasing magnitude of $m'$ for a particular $S$ while the charge conductance  for a particular $S$ varies with the value of magnetic moment but is independent of the sign. Further, for increasing magnitude of $m'$, $G_{c}$ increases. So to characterize the ground state spin and magnetic moment of the high spin magnetic impurity we look at Table 1 and 2 in conjunction, we first determine spin of magnetic impurity $S$ from $G_c$ values of Table 1 then we determine magnitude of magnetic moment $m'$ from $G_c$ values of Table 2, finally to determine the sign  of magnetic moment we look at $G_s$ values of either Table 1 or Table 2. We thus can characterize the total ground state spin and magnetic moment of the high spin magnetic impurity.  In tables 1 and 2 we also feature the  spin flip probabilities for magnetic impurity- $F_2(=\sqrt{(S-m')(S+m'+1)})$ when spin up electron is incident\cite{A,AJP} while $F_4(=\sqrt{(S+m')(S-m'+1)}$ is when spin down electron is incident.

In order to get a closed form empirical expression for the differential charge and  spin conductance  we first plot the differential charge conductance as function of the spin of magnetic impurity $S$ for $E=0$, $E=0.1\Delta$, $E=0.5\Delta$, $E=\Delta$ limit for transparent ($Z=0$) case in Fig.~6.
\subsection{Characterizing the spin S of HSM}

A least square fit in Figs.~6(a-c) indicates that the charge conductance exponentially decays with the spin ($S$) of the magnetic impurity, thus $G_{c} \propto exp(-S)$. To plot the Figs.~6(a-c)  we take the mean charge conductance for a particular $S$, the reason being although $G_c$ is fairly constant for different values of $m'$ there is a slight increase for increasing magnitude of $m'$.
\begin{figure}[h]
\includegraphics[width=.84\textwidth]{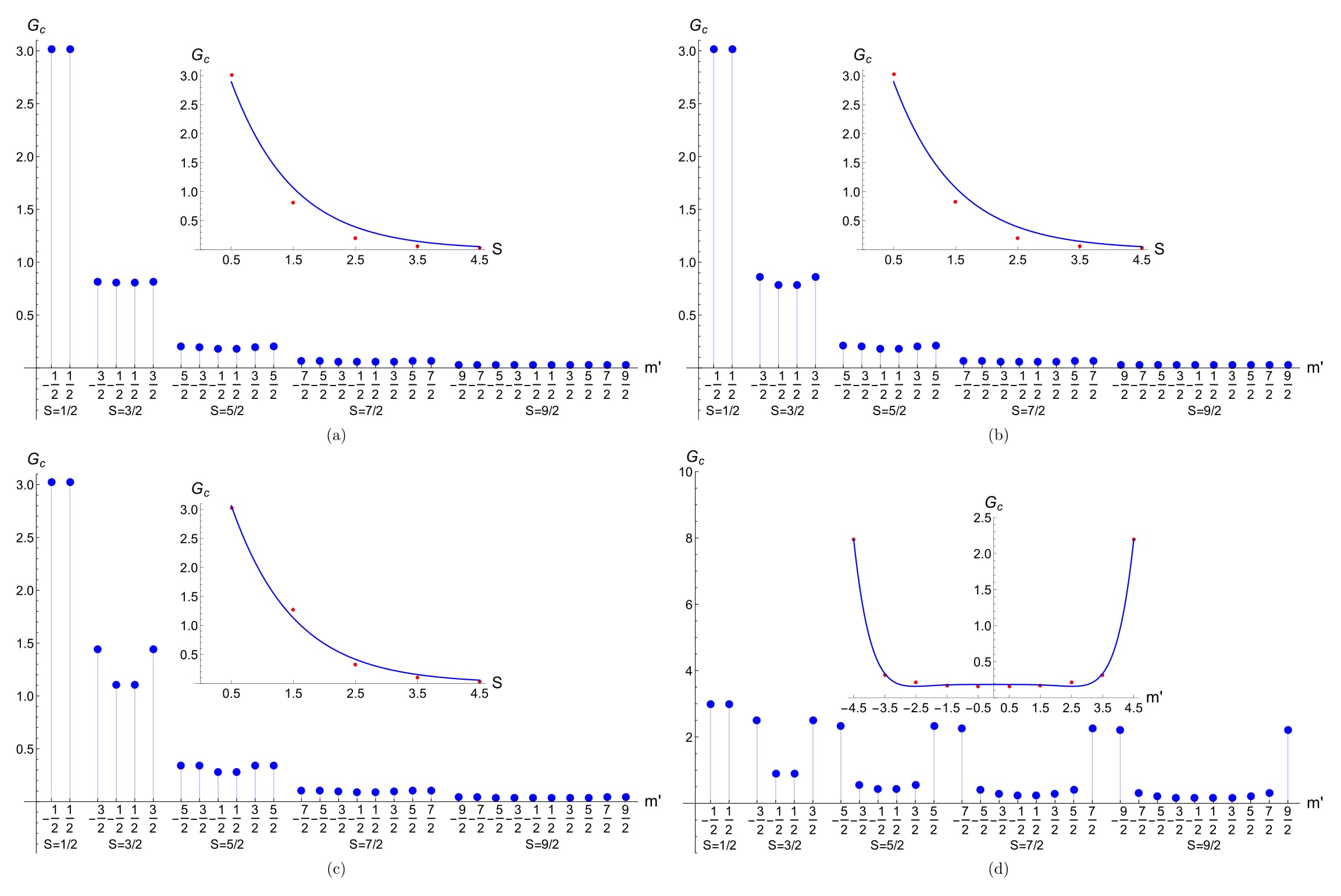}
\caption{ a) \small \sl Differential  charge conductance vs impurity spin plot for $E=0, J=1, Z=0$. Here fitting function is $4.76995 e^{-S}$ b)Charge conductance vs impurity spin plot for $E=0.1\Delta,J=1,Z=0$. Here fitting function is $4.77961 e^{-S}$ c) \small \sl Charge conductance vs impurity spin plot for $E=0.5\Delta,J=1,Z=0$. Here fitting function is $5.05004 e^{-S}$  d) Charge conductance vs impurity spin plot for $E=\Delta,J=1$. Here fitting function is $0.183844-0.00123696m^{\prime 4}+0.0000149948m^{\prime 8}$ in d) the plot is independent of whether we are in transparent or tunneling regime.}
\end{figure}
One can also see the charge conductance does not change dramatically from the zero energy limit as is evident from Fig.~6(b), $E=0.1\Delta$ and 6(c) $E=0.5\Delta$.  This approximate relation $G_{c}\propto exp(-S)$ for the charge conductance implies that one can exactly characterize the spin (S) of the magnetic impurity irrespective of the voltage applied at the metal superconductor interface as long as we are below the gap. The situation changes at and near the gap edge $E=\Delta$, shown in Fig.~6(d), herein the charge conductance almost does not vary with S, and the variation with $m'$ for different S is also uniform.
\subsection{Characterizing the magnetic moment $m'$ of HSM}
Next we plot the spin conductance again for transparent case ($Z=0$) in Fig.~7 and for the tunneling regime in Fig.~8. Unlike the charge conductance the spin conductance does not show any definite pattern as function of S, apart from the fact that as S increases there is a monotonic decrease in spin conductance. However, an interesting pattern emerges when the spin conductance is plotted for a particular S as function of the magnetic moment $m'$, in this case as one can make out that a least square fit for both transparent case ($Z=0$) and the tunneling regime take the general form $G_{s}\propto -m'$.
\begin{figure}[h]
\includegraphics[width=.9\textwidth]{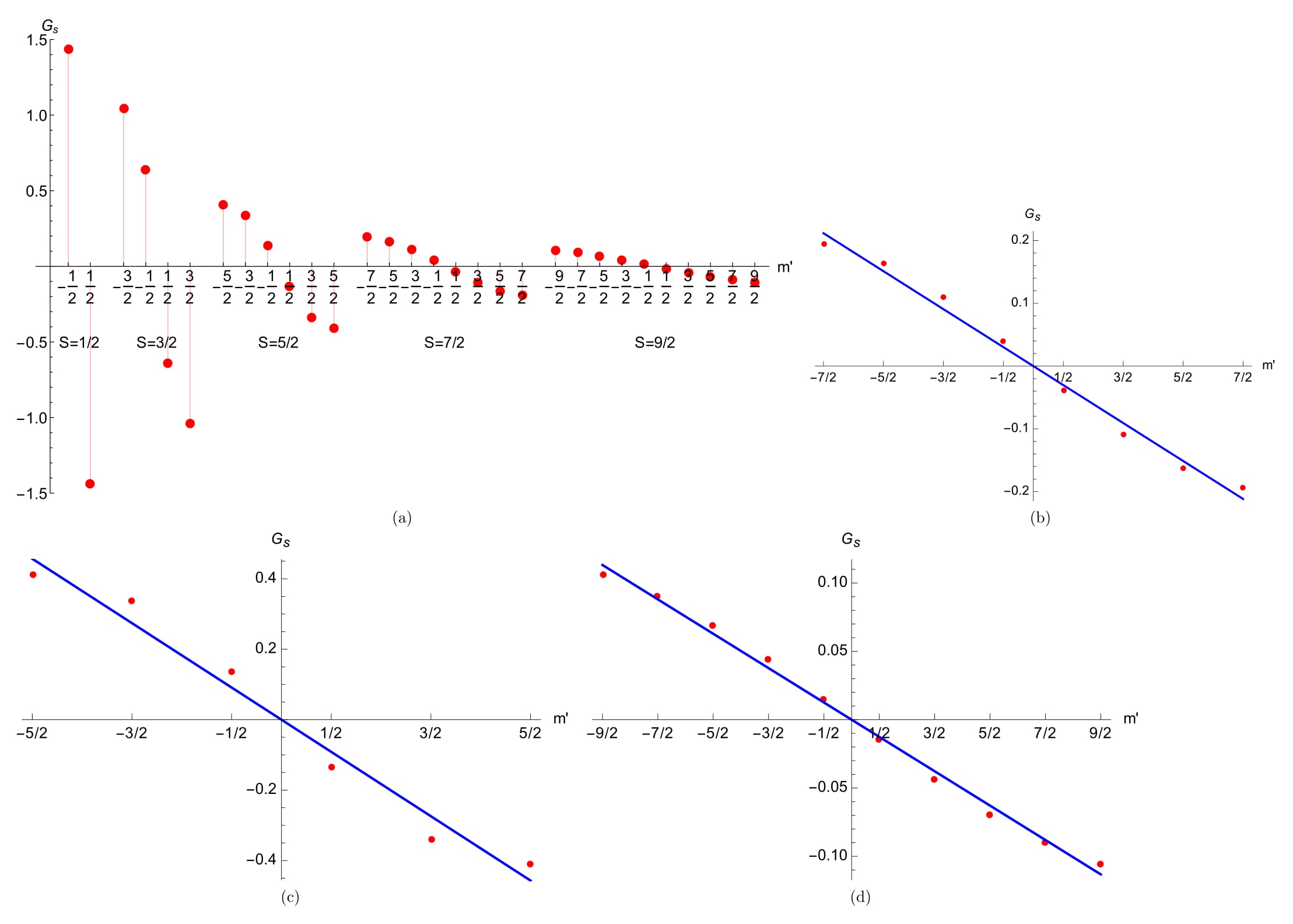}
\caption{ a) \small \sl Differential spin conductance vs impurity spin plot for $E=0, J=1, Z=0$ transparent case, b)least square fit plot for $S=7/2, E=0, J=1, Z=0$. Here fitting function is $-0.0604876 m'$c)least square fit plot for $S=5/2,E=0, J=1, Z=0$. Here fitting function is $-0.182785 m'$ d)least square fit plot for $S=9/2,E=0,J=1,Z=0$. Here fitting function is $-0.0251511 m'$ }
\end{figure}
\begin{figure}[h]  
\includegraphics[width=.8\textwidth]{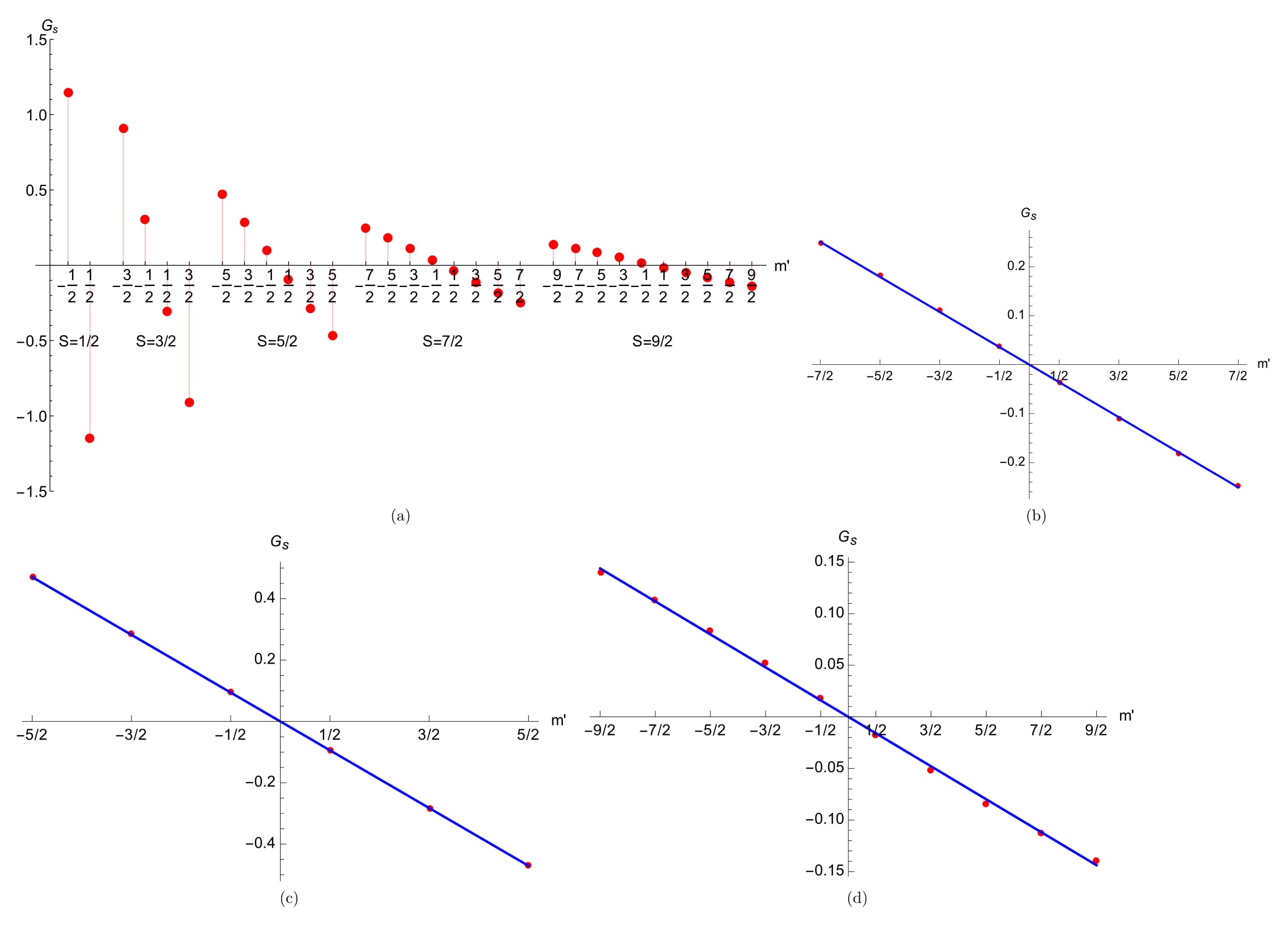}
\caption{ a) \small \sl Spin conductance vs impurity spin plot for $E=0, J=1, Z=3$ tunneling case, b)least square fit plot for $S=7/2,E= 0,J=1,Z=3$. Here fitting function is $- 0.071622 m'$c)least square fit plot for $S=5/2,E=0, J=1, Z=3$. Here fitting function is $- 0.188307 m'$ d)least square fit plot for $S=9/2,E= 0, J=1, Z=3$. Here fitting function is $- 0.0319619 m'$ }
\end{figure}
\subsection{Characterizing the magnitude and nature of exchange interaction between HSM and electron/hole in the system}
In Fig.~9 we have plotted charge conductance as a function of J for a high spin magnetic impurity (S=5/2). We see that as the magnetic moment ($m'$) increases charge conductance increases. The variation of $G_{c}$ with J is Lorentzian. The least square function fits are plotted too, these are for plot Fig 9(b) ($E=0,Z=0,S=5/2$) the fit function is $ G_{c}= 1.87056/(1+11.5149J^{2}) + 2.16776 \exp(-4.35263J^{2})$, in the E=0 limit the charge conductance is almost independent of magnetic moment $m'$.
For Fig. 9(a) ($E=\Delta,Z=0, S=5/2$) for $m'=5/2$, fit function is $G_{c}=2+2/(1+5J^{2})$,  for  $m'=3/2$ fit function is $G_{c}= 4.10204/(1+6.36483J^{2}) - 0.104924 \exp(-5.17046J^{2})$ and finally for $m'=1/2$ fit function is $G_{c}= 4.00646/(1+8.48884J^{2}) - 0.0066353 \exp(-6.96064J^{2})$. This helps in estimating the strength of interaction, however the sign of interaction remains undecipherable from this plot.  To estimate the sign of interaction we plot the derivative of differential spin conductance ($dG_{s}/dJ$) with respect to $J$ in Fig. 9(c). We can clearly see that this quantity is anti-symmetric with respect to  J, enabling us to detect the nature of interaction of electrons/holes in the system with the HSM.
\begin{figure}[h]  
\includegraphics[width=.83\textwidth]{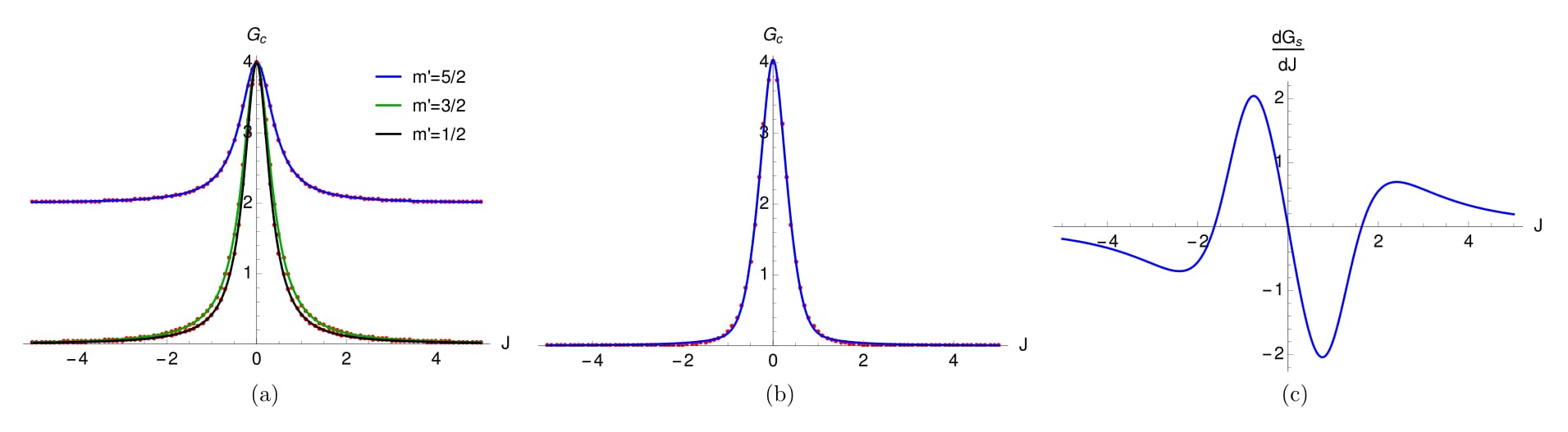}
\caption{ a) \small \sl Charge conductance vs. exchange interaction strength J for(a) $S=5/2$ and different values of magnetic moment of magnetic impurity. Other parameters are $Z=0, \mbox  { and }  E=\Delta$. (b)  $S=5/2, E=0, Z=0$. For this case differential charge conductance becomes almost independent of $m'$ and (c) The derivative of differential spin conductance with respect to $J$. Regardless of $ S, m' $, $\frac{dGs}{dJ}$ is antisymmetric with respect to the nature of interaction. From this plot we can note the sign of the interaction.}
\end{figure}
\section{Case of finite temperature}
According to the BTK theory, the charge current at temperature $T$ subjected to bias voltage $V$ is given by
\begin{equation}
I_{c}=\frac{2N(0)ev_{F}A}{2\pi}\int_{0}^{2\pi}\int_{-\infty}^{\infty}[f_{0}(E-eV)-f_{0}(E)]G_{c}^{0}dEd(k_{F}a) 
\end{equation}
Similarly, the spin current at temperature $T$ subjected to bias voltage $V$ is given by
\begin{equation}
I_{s}=\frac{2N(0)ev_{F}A}{2\pi}\int_{0}^{2\pi}\int_{-\infty}^{\infty}[f_{0}(E-eV)-f_{0}(E)]G_{s}^{0}dEd(k_{F}a)  
\end{equation}
where $A=\pi a^2/4$ is an effective-neck cross-sectional area, including a numerical factor for angular averaging which will depend on the actual
$3D$ geometry and $a$ is the radius of the orifice. $N(0)$ denotes the one-spin density of states at $E_{F}$ and $v_{F}$ is the Fermi velocity.
In Fig.~11 we have plotted charge and spin current as a function of exchange interaction strength $J$ for finite temperature. In Fig.~11(a) we plot charge current, here we take $T=1K, S=5/2$ and for all possible values of magnetic moments. Here with increase of magnetic moment charge current decreases and it is independent on the sign of magnetic moment. In Fig.~11(b) we plot charge current with taking $T=0.2K$.
We see that with decrease of temperature charge current increases, but the nature of the plot remains unchanged. Similarly in Fig.~11(c) and (d) we plot spin current as a function of exchange interaction strength $J$ for $T=1K$ and $T=0.2K$ respectively and other parameters remain same. Here also with decrease of temperature spin current increases, but it increases with the magnitude of $m'$.
\begin{figure}[h]  
\includegraphics[width=.7\textwidth]{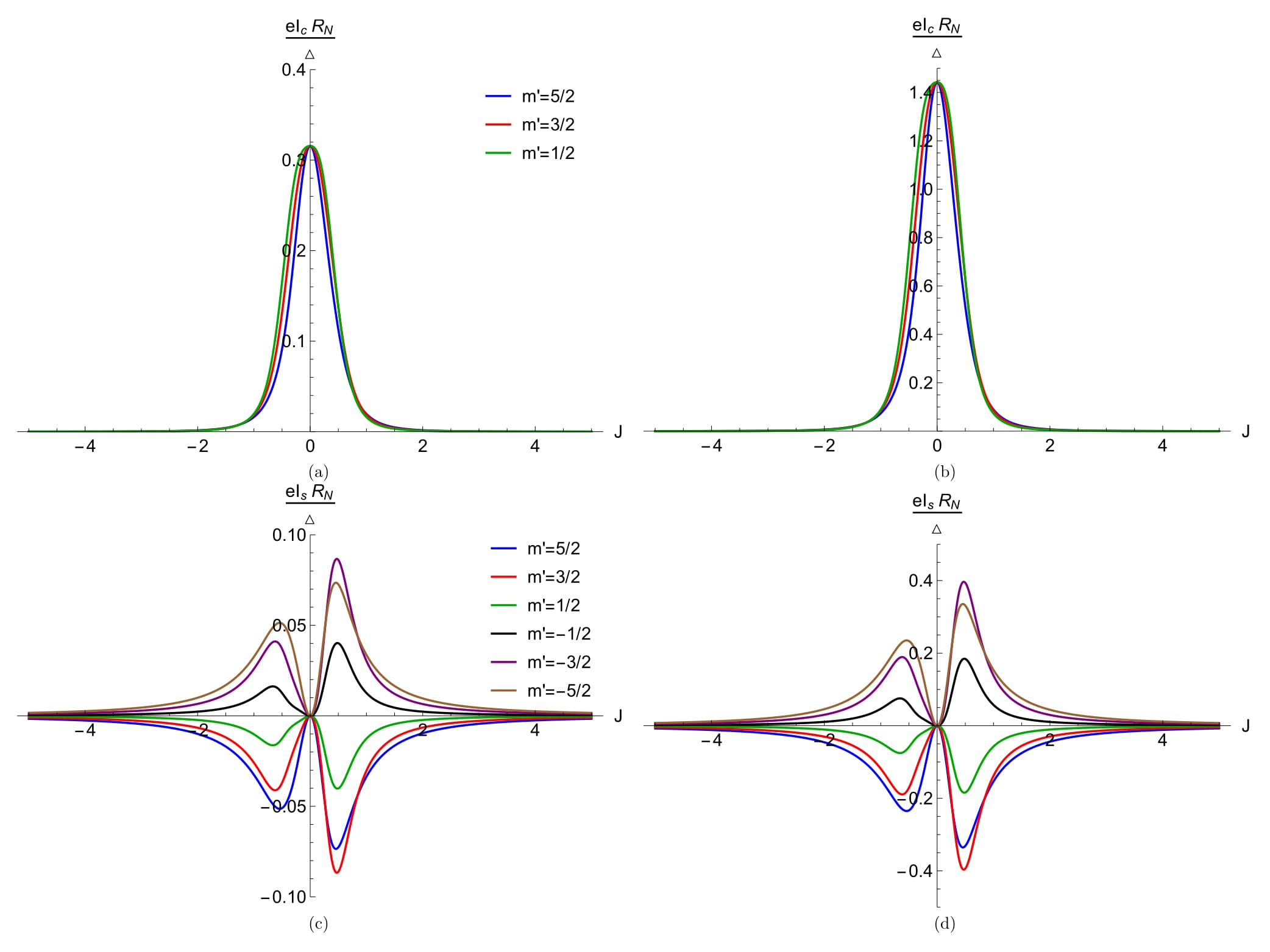}
\caption{Finite temperature plots for a) \small \sl Charge current vs. exchange interaction strength J with $T=1K$ and different values of magnetic moment of magnetic impurity (b) Charge current vs. exchange interaction strength J with $T=0.2K$ and different values of magnetic moment of magnetic impurity (c) Spin current vs. exchange interaction strength J with $T=1K$ and different values of magnetic moment of magnetic impurity (d) Spin current vs. exchange interaction strength J with $T=0.2K$ and different values of magnetic moment of magnetic impurity.Other parameters are $S=5/2,Z=0,eV=\Delta/2$ and $R_{N}=(1+Z^2)/2N(0)e^2v_{F}A$.}
\end{figure}
\section{Application to spintronics: Pure spin conductance in absence of charge }
 Next in Fig.~12 we compare the differential spin and charge conductance for NIS, FIS and NMNIS junctions in the transparent limit(Z=0). Further, the polarization in Ferromagnet is $90\%$ (as is indicated in (a) $h_{0}=0.9 E_{F}$).
 The parameters for NMNIS are mentioned in Figure itself. In Fig.~12(a) the differential spin conductance is plotted, it shows remarkably that there is a finite spin conductance below the gap in case of a NMNIS junction in contrast to either a FIS or NIS junction. In Fig.~12(b), we plot the charge conductance ($G_{c}$), while in the NIS junction $G_{c}$ is constant at the value $4$. On the other hand for a FIS junction, it is always less than $4$ and beyond a peak at $E=0$, $G_{c}$ continuously decreases. In the NMNIS junction, $G_{c}$ is almost constant at the value $3$. Further, in Fig.~13 (a) and (b) we plot $G_{c}$ and $G_{s}$ along-with the individual up-spin ($G_{s}^{\uparrow}$) and down-spin ($G_{s}^{\downarrow}$) conductances.
 We see in Fig.~13(b) that pure spin conductance with only selective transport of spin-up without any charge transport while in Fig.~13(a) it is shown that pure spin conductance with only selective transport of spin-down without any charge transport.
\begin{figure}[h]  
\includegraphics[width=.67\textwidth]{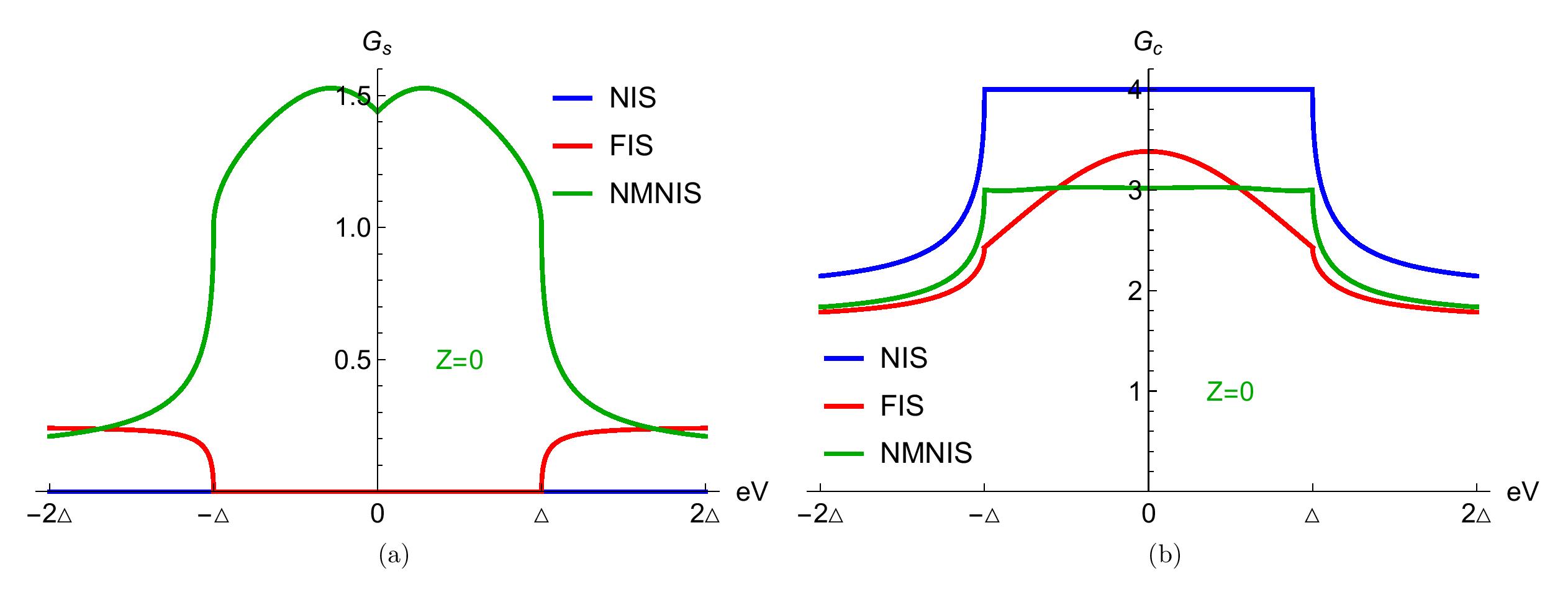}
\caption{ a) \small \sl Spin conductance and (b) Charge conductance for NIS, FIS and NMNIS junctions in the transparent regime. In the FIS junction, magnetization is $h_{0}=0.9 E_{F}$. In the NMNIS junction, spin and magnetic moment of magnetic impurity $S=1/2, m'=-1/2$ and $J=1$. Note the absence of any spin conductance below the gap for NIS and FIS junction.}
\end{figure}
Finally, in Fig.~14(a) and (b) the charge and spin conductances are plotted for HSM with $S=1/2$ and $m'=\pm 1/2$ as function of the transparency of the junction $Z$. The charge conductance is independent of the sign of spin magnetic moment $m'$ but the spin conductance  is dependent and it has to be noted that the spin conductance has same magnitude but opposite sign for different sign of $m'$. To check what happens when we increase $S$ we plot $G_{c}$ and $G_{s}$ as function of $Z$ in Fig.~14(c) and (d) for $S=5/2$ and different values of possible spin magnetic moment $m'=-5/2,-3/2,-1/2,1/2,3/2,5/2$. $G_{c}$ being independent of sign of $m'$ but depends on the magnitude of $m'$. One thing which is quite apparent is that as Z increases into the tunneling regime greater than $3$, $G_{c}\rightarrow 0$. $G_{s}$ on the other hand in the tunneling regime is constant and finite and the magnitude of spin conductance increases with the magnitude of spin magnetic moment.
\begin{figure}[h]  
\includegraphics[width=.8\textwidth]{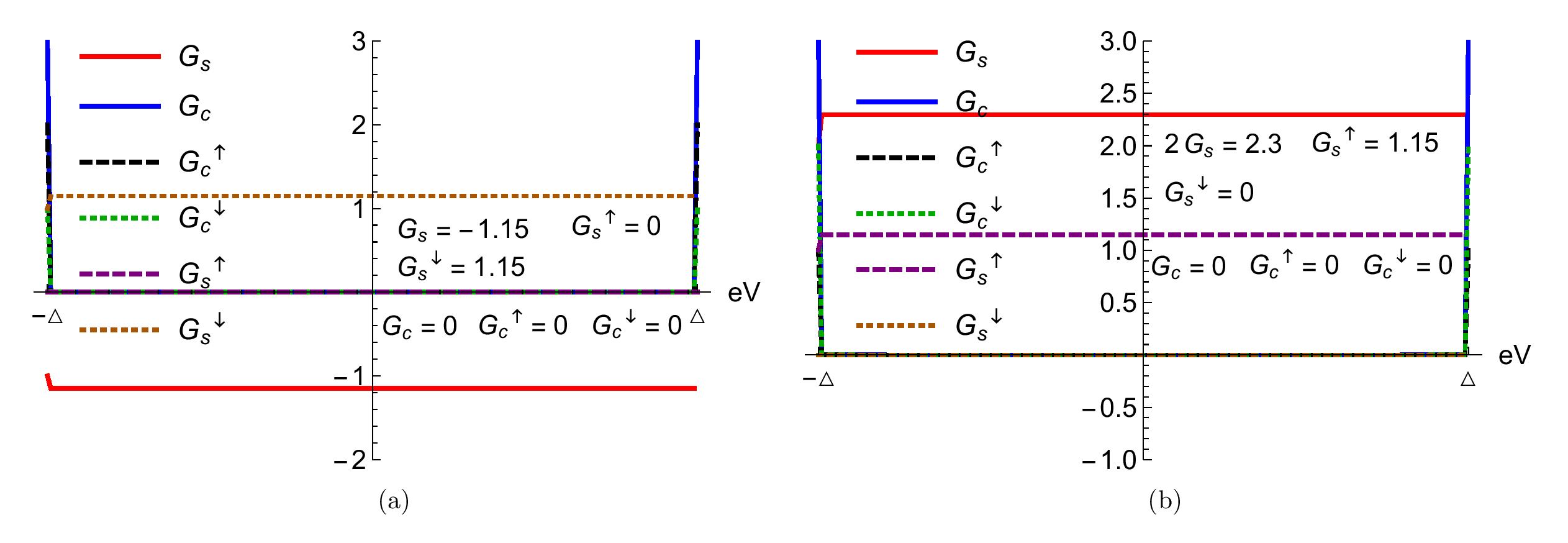}
\caption{Tunneling regime plots for a) \small \sl Spin conductance and Charge conductance vs energy with $J=1,Z=10,S=1/2,m'=1/2$ b) Spin conductance and Charge conductance vs energy with $J=1,Z=10,S=1/2,m'=-1/2$. Note the exclusive pure spin conductance (only spin down electrons contribute) in absence of any charge for (a) and exclusive pure spin conductance (only spin up electrons contribute) in absence of any charge for (b).}
\end{figure}
\begin{figure}[h]
\includegraphics[width=.75\textwidth]{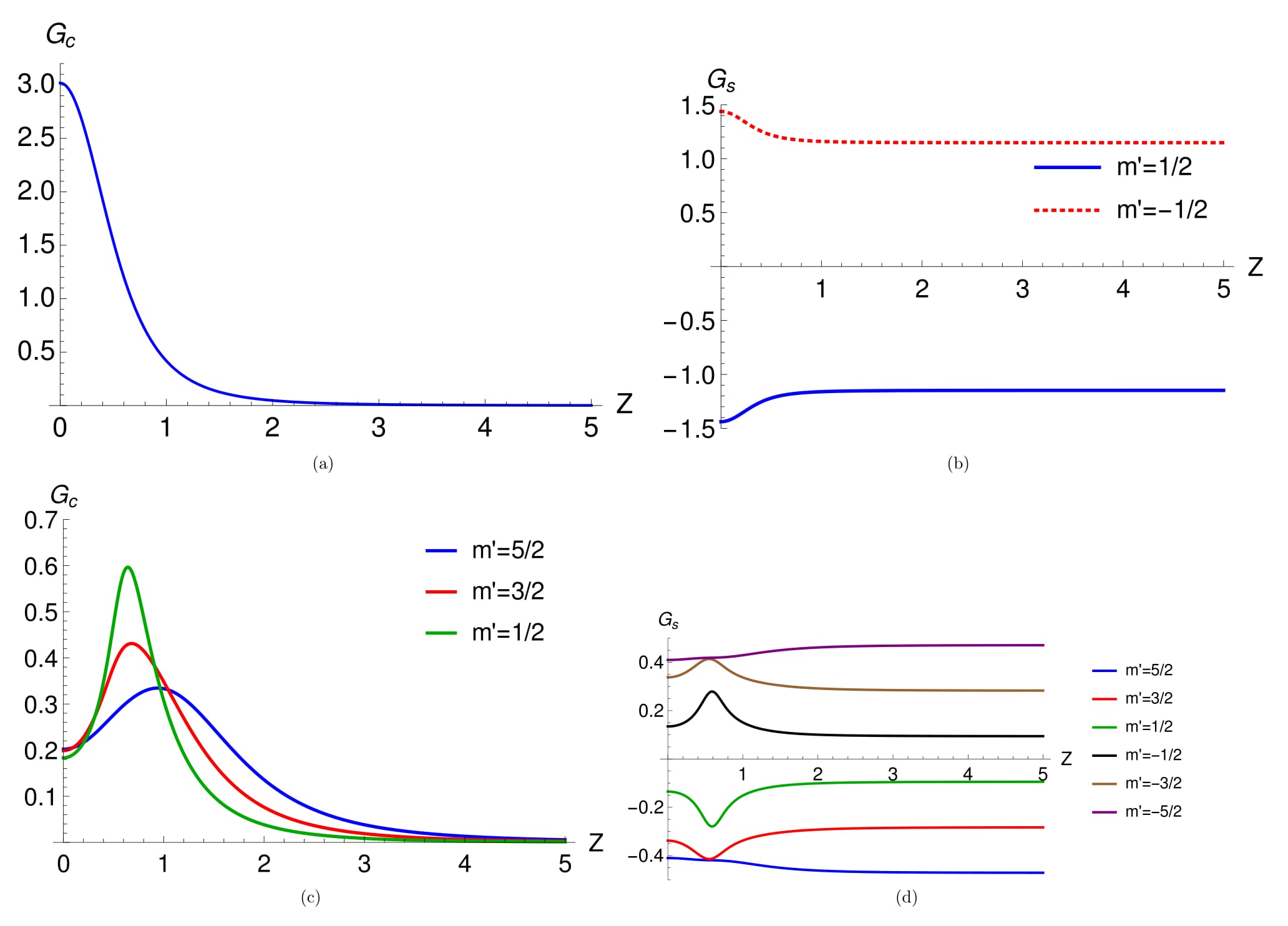}
\caption{a) \small \sl Charge conductance vs Z plot for $J=1,S=1/2,E=0, m'=1/2$ b)Spin conductance vs Z plot for $J=1,S=1/2,E=0$ c)Charge conductance vs Z plot for $J=1,S=5/2,E=0$ d)Spin conductance vs Z plot for $J=1,S=5/2,E=0.$}
\end{figure}
\newpage
\section {Conclusions and Experimental Realization}
To conclude, we aimed to fulfill two objectives with our set-up as outlined in the Introduction. The first of fully characterizing a HSM was accomplished using the charge and spin conductance of the NMNIS junction at the zero bias limit $E=0$ and the gap edge limit $E=\Delta$. The second aim of pure spin conductance without any accompanying charge conductance was realized in the tunneling $Z \rightarrow large$ limit. 
The set-up as envisaged in Fig.~1(a) and (b), can be easily realized in the lab. The NIS junctions have been experimentally realized since more than 30 years\cite{Gus}. High spin magnetic impurities have been realized since 20 years. The amalgamation of a NIS junction with a HSM shouldn't be difficult. Especially with a s-wave superconductor like Aluminium or Niobium it should be perfectly possible. In future junctions with HSM and High T$_{c}$ superconductors will be attempted, this will have the additional aspect of nodes in the superconducting gap, and the aim would be to exploit it for spintronics. 
\section{Appendix} The Appendix consists of the details of wavefunctions and boundary conditions needed to solve the scattering problem, also with two Tables I and II needed to understand some of the results of the work accompanies this manuscript.
In section I we first introduce the wave functions of the different region of the system. We outline a brief sketch of our set-up and and provide a theoretical background to our study by writing the Hamiltonian and solving the equations after imposing boundary conditions on the wavefunctions. In section II and III we give two tables where charge and spin conductance values are shown for different values of spin ($S$) and  magnetic moment ($m'$) of the magnetic impurity in the zero bias limit ($E=0$) and the gap edge limit ($E=\Delta$) respectively. To characterize the ground state spin and magnetic moment of the high spin magnetic impurity we look at Table I and II in conjunction, we first determine spin of magnetic impurity $S$ from $G_{c}$ values of Table I then we determine magnitude of magnetic moment $m'$ from $G_{c}$ values of Table II, finally to determine the sign of magnetic moment we look at $G_{s}$ values of either Table I or Table II. We thus can characterize the total ground state spin and magnetic moment of the high spin magnetic impurity.
\subsection{Wavefunctions and boundary conditions in the metal-magnetic impurity-metal-insulator-superconductor junction}
The wave functions of the different region of the system as shown in FIG. 1(a) and FIG. 1(b) can be written in spinorial form\cite{LINDER}:
\[\psi_{N}^{I}(x)=\begin{bmatrix}
                    1\\
                    0\\
                    0\\
                    0
                  \end{bmatrix}e^{ik_{e}x}\phi_{m'}^{S}+r_{ee}^{\uparrow\uparrow}\begin{bmatrix}
                  1\\
                  0\\
                  0\\
                  0
                 \end{bmatrix}e^{-ik_{e}x}\phi_{m'}^{S}+r_{ee}^{\uparrow\downarrow}\begin{bmatrix}
                 0\\
                 1\\
                 0\\
                 0
                \end{bmatrix}e^{-ik_{e}x}\phi_{m'+1}^{S}+r_{eh}^{\uparrow\uparrow}\begin{bmatrix}
                0\\
                0\\
                1\\
                0
               \end{bmatrix}e^{ik_{h}x}\phi_{m'+1}^{S}+r_{eh}^{\uparrow\downarrow}\begin{bmatrix}
               0\\
               0\\
               0\\
               1
              \end{bmatrix}e^{ik_{h}x}\phi_{m'}^{S},\mbox{for $x<0$}\]
\begin{eqnarray}
\Psi_{N}^{II}(x)=t_{ee}^{'\uparrow\uparrow}\begin{bmatrix}
                                     1\\
                                     0\\
                                     0\\
                                     0
                                     \end{bmatrix}e^{ik_{e}x}\phi_{m'}^{S}+t_{ee}^{'\uparrow\downarrow}\begin{bmatrix}
                                     0\\
                                     1\\
                                     0\\
                                     0
                                    \end{bmatrix}e^{ik_{e}x}\phi_{m'+1}^{S}+b_{ee}^{\uparrow\uparrow}\begin{bmatrix}
                                    1\\
                                    0\\
                                    0\\
                                    0
                                   \end{bmatrix}e^{-ik_{e}(x-a)}\phi_{m'}^{S}+b_{ee}^{\uparrow\downarrow}\begin{bmatrix}
                                   0\\
                                   1\\
                                   0\\
                                   0
                                  \end{bmatrix}e^{-ik_{e}(x-a)}\phi_{m'+1}^{S}\nonumber\\+c_{eh}^{\uparrow\uparrow}\begin{bmatrix}
                                  0\\
                                  0\\
                                  1\\
                                  0
                                 \end{bmatrix}e^{ik_{h}(x-a)}\phi_{m'+1}^{S}+c_{eh}^{\uparrow\downarrow}\begin{bmatrix}
                                 0\\
                                 0\\
                                 0\\
                                 1
                                \end{bmatrix}e^{ik_{h}(x-a)}\phi_{m'}^{S}+a_{eh}^{\uparrow\uparrow}\begin{bmatrix}
                                0\\
                                0\\
                                1\\
                                0
                               \end{bmatrix}e^{-ik_{h}x}\phi_{m'+1}^{S}+a_{eh}^{\uparrow\downarrow}\begin{bmatrix}
                               0\\
                               0\\
                               0\\
                               1
                              \end{bmatrix}e^{-ik_{h}x}\phi_{m'}^{S},\mbox{for $0<x<a$}\nonumber
                              \end{eqnarray}
\[\psi_{S}(x)=t_{ee}^{\uparrow\uparrow}\begin{bmatrix}
                              u\\
                              0\\
                              0\\
                              v
                             \end{bmatrix}e^{iq_{+}x}\phi_{m'}^{S}+t_{ee}^{\uparrow\downarrow}\begin{bmatrix}
                             0\\
                             u\\
                             -v\\
                             0
                             \end{bmatrix}e^{iq_{+}x}\phi_{m'+1}^{S}+t_{eh}^{\uparrow\uparrow}\begin{bmatrix}
                             0\\
                             -v\\
                             u\\
                             0
                             \end{bmatrix}e^{-iq_{-}x}\phi_{m'+1}^{S}+t_{eh}^{\uparrow\downarrow}\begin{bmatrix}
                             v\\
                             0\\
                             0\\
                             u
                             \end{bmatrix}e^{-iq_{-}x}\phi_{m'}^{S},\mbox{for $x>a$}\] \\
 $r_{ee}^{\uparrow\uparrow}$($r_{ee}^{\uparrow\downarrow}$) and $r_{eh}^{\uparrow\uparrow}$($r_{eh}^{\uparrow\downarrow}$) are the corresponding  amplitudes for normal reflection and Andreev reflection with spin up(down). $t_{ee}^{\uparrow\uparrow}$($t_{ee}^{\uparrow\downarrow}$) and $t_{eh}^{\uparrow\uparrow}$($t_{eh}^{\uparrow\downarrow}$) are the corresponding amplitudes for transmission of electron-like quasi-particles and hole-like quasi-particles with spin up(down). $\phi_{m'}^{S}$ is the eigenfunction of magnetic impurity: with its $S^{z}$ operator acting as- $S^{z}\phi_{m'}^{S} = m'\phi_{m'}^{S}$, with $m'$ being the spin magnetic moment of the HSM. 
For $E>\Delta$(for energies above the gap),the coherence factors are $u^2=\frac{1}{2}\Big[\frac{E+(E^2-\Delta^2)^\frac{1}{2}}{E}\Big]$, $v^2=\frac{1}{2}\Big[\frac{E-(E^2-\Delta^2)^\frac{1}{2}}{E}\Big]$, while the wave-vector in metal is $k_{e,h}=\sqrt{2m^{\star}(E_{F}\pm E)}$ and in superconductor is $q_{\pm}=\sqrt{2m^{\star}(E_{F}\pm \sqrt{E^2-\Delta^2})}$ and for $E<\Delta$(for energies below the gap) the coherence factors are {$u^2=\frac{1}{2}\Big[\frac{E+i(\Delta^2-E^2)^\frac{1}{2}}{\Delta}\Big]$, $v^2=\frac{1}{2}\Big[\frac{E-i(\Delta^2-E^2)^\frac{1}{2}}{\Delta}\Big]$}, while the wavevector in metal is 
$k_{e,h}=\sqrt{2m^{\star}(E_{F}\pm E)}$ and in superconductor is $q_{\pm}=\sqrt{2m^{\star}(E_{F}\pm i\sqrt{\Delta^2-E^2})}$\cite{BTK}, wherein $E_{F}$ is the Fermi energy, $m^{*}$ is the effective mass of electron in metal and $E$ is the excitation energy of electron above $E_{F}$. In Andreev approximation, which we will use throughout this work, $E_{F}\gg \Delta, E$ we take $k_{e}=k_{h}=q_{+}=q_{-}=k_{F}$.
We impose the boundary conditions on the above wave-functions and solve the resulting $16$ equations and get the different scattering amplitudes: $r_{ee}^{\uparrow\uparrow},r_{ee}^{\uparrow\downarrow},r_{eh}^{\uparrow\uparrow},r_{eh}^{\uparrow\downarrow},t_{ee}^{\uparrow\uparrow},t_{ee}^{\uparrow\downarrow},t_{eh}^{\uparrow\uparrow},t_{eh}^{\uparrow\downarrow}$. The reflection and transmission probabilities we get are thus-  $R_{ee}^{\uparrow\uparrow}= |r_{ee}^{\uparrow\uparrow}|^{2},R_{ee}^{\uparrow\downarrow}=|r_{ee}^{\uparrow\downarrow}|^{2},R_{eh}^{\uparrow\uparrow}=|r_{eh}^{\uparrow\uparrow}|^{2},R_{eh}^{\uparrow\downarrow}=|r_{eh}^{\uparrow\downarrow}|^{2}$, {$T_{ee}^{\uparrow\uparrow}=(u^2-v^2)|t_{ee}^{\uparrow\uparrow}|^{2}, T_{ee}^{\uparrow\downarrow}=(u^2-v^2)|t_{ee}^{\uparrow\downarrow}|^{2},T_{eh}^{\uparrow\uparrow}=(u^2-v^2)|t_{eh}^{\uparrow\uparrow}|^{2},T_{eh}^{\uparrow\downarrow}=(u^2-v^2)|t_{eh}^{\uparrow\downarrow}|^{2}$}.\\
We consider a metal (N)-metal (N)-superconductor (S) junction where there is a HSM between two metallic regions at $(x=0)$ and a $\delta$-like potential barrier exists at metal superconductor interface at $(x=a)$. When an electron with energy E and spin ($\uparrow/\downarrow$) is incident from the normal metal, at the $x=0$ interface it interacts with the HSM through an exchange potential which may induce a mutual spin flip. The electron can be reflected back to region I, or transmitted to region II, with spin up or down. When this transmitted electron is incident at $x=a$ interface it could be reflected back from the interface and there is also possibility of Andreev reflection, i.e., a hole with spin up or down is reflected back to region II. Electron-like and hole-like quasiparticles with spin up or down are transmitted into the superconductor for energies above the gap.
The model Hamiltonian in Bogoliubov-de Gennes formalism of our Normal Metal-Magnetic impurity-Normal Metal-Insulator-Superconductor system is given below:
\begin{eqnarray}
  \begin{bmatrix}
    H\hat{I} & i\Delta \theta(x-a)\hat{\sigma}_{y}  \\
    -i\Delta^{*}\theta(x-a)\hat{\sigma}_{y}   &  -H\hat{I}
  \end{bmatrix} \Psi(x)& =& E \Psi(x), \mbox{ with } H = p^2/2m^\star + V\delta(x-a) - J_{0}\delta(x)\vec s.\vec S -E_{F},\nonumber\\
&&\mbox{ $\Psi$ is a four-component spinor, $\Delta$ is the gap in s-wave superconductor and }\nonumber\\
&&\theta \mbox{   is the Heaviside step function.}
\end{eqnarray}
Further, in $H$ the first term is the kinetic energy of an electron with effective mass $m^\star$, for second term  $V$ is the strength of the $\delta$-like potential at the interface between normal metal and superconductor, the third term describes the exchange interaction of strength $J_{0}$ between the electron with spin $\vec s$ and a magnetic impurity with spin $\vec S$, $\hat{\sigma}$ is the Pauli spin matrix and $\hat{I}$ is unit matrix, $E_F$ being the Fermi energy. We will later use the dimensionless parameter $J=\frac{m^{\star}J_{0}}{\hbar^2 k_{F}}$ as a measure of strength of exchange interaction\cite{AJP} and $Z=\frac{m^{\star}V}{\hbar^2 k_{F}}$ as a measure of interface transparency\cite{BTK}. In our work $Z$ is a dimensionless quantity, while $V$ has the dimension of energy. $Z$ denotes the transparency of the junction, $Z=0$ means completely transparent junction, while $Z>>1$ implies a tunneling junction\cite{BTK,b}.\\ 
\begin{figure}[h]
\centering{\includegraphics[width=.7\linewidth]{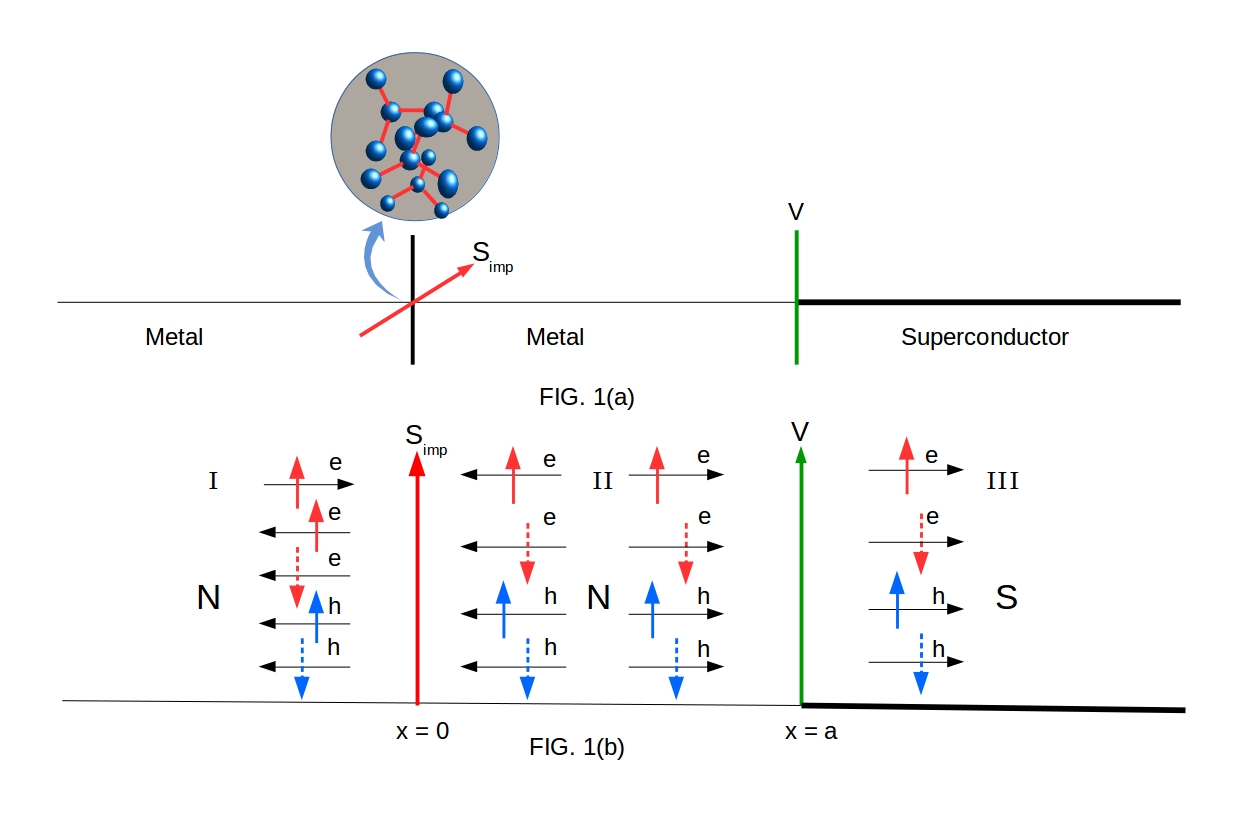}}
\caption{(a) A high spin magnetic impurity(HSM) with spin S and magnetic moment $m'$ at $x=0$ in a Normal Metal-HSM-Normal Metal-Insulator-Superconductor (NMNIS) junction, (b) The scattering of an up-spin electron incident is shown. Andreev reflection and quasi particle transmission into superconductor are depicted.}
\end{figure}
At $x=0$
\begin{equation}
\psi_{N}^{I}(x)=\psi_{N}^{II}(x),\mbox{(continuity of wavefunctions)}
\end{equation}
\begin{equation}
\frac{d\psi_{N}^{II}}{dx}-\frac{d\psi_{N}^{I}}{dx}=-\frac{2m^{\star}J_{0}\vec s.\vec S}{\hbar^2} \psi_{N}^{I},\mbox{(discontinuity in first derivative)} 
\end{equation}
At $x=a$
\begin{equation}
 \psi_{N}^{II}(x)=\psi_{S}(x),\mbox{(continuity of wavefunctions)}
\end{equation}
\begin{equation}
\frac{d\psi_{S}}{dx}-\frac{d\psi_{N}^{II}}{dx}=\frac{2m^{\star}V}{\hbar^2}\psi_{N}^{II},\mbox{(discontinuity in first derivative)} 
\end{equation}
From boundary conditions $(14)$ and $(16)$ we get
\begin{equation}
1+r_{ee}^{\uparrow\uparrow}=t_{ee}^{'\uparrow\uparrow}+b_{ee}^{\uparrow\uparrow}e^{ik_{e}a} 
\end{equation}
\begin{equation}
r_{ee}^{\uparrow\downarrow}=t_{ee}^{'\uparrow\downarrow}+b_{ee}^{\uparrow\downarrow}e^{ik_{e}a} 
\end{equation}
\begin{equation}
r_{eh}^{\uparrow\uparrow}=c_{eh}^{\uparrow\uparrow}e^{-ik_{h}a}+a_{eh}^{\uparrow\uparrow}
\end{equation}
\begin{equation}
r_{eh}^{\uparrow\downarrow}=c_{eh}^{\uparrow\downarrow}e^{-ik_{h}a}+a_{eh}^{\uparrow\downarrow} 
\end{equation}
\begin{equation}
t_{ee}^{'\uparrow\uparrow}e^{ik_{e}a}+b_{ee}^{\uparrow\uparrow}=t_{ee}^{\uparrow\uparrow}ue^{iq_{+}a}+t_{eh}^{\uparrow\downarrow}ve^{-iq_{-}a} 
\end{equation}
\begin{equation}
t_{ee}^{'\uparrow\downarrow}e^{ik_{e}a}+b_{ee}^{\uparrow\downarrow}=t_{ee}^{\uparrow\downarrow}ue^{iq_{+}a}-t_{eh}^{\uparrow\uparrow}ve^{-iq_{-}a}
\end{equation}
\begin{equation}
c_{eh}^{\uparrow\uparrow}+a_{eh}^{\uparrow\uparrow}e^{-ik_{h}a}=t_{eh}^{\uparrow\uparrow}ue^{-iq_{-}a}-t_{ee}^{\uparrow\downarrow}ve^{iq_{+}a}
\end{equation}
\begin{equation}
c_{eh}^{\uparrow\downarrow}+a_{eh}^{\uparrow\downarrow}e^{-ik_{h}a}=t_{eh}^{\uparrow\downarrow}ue^{-iq_{-}a}+t_{ee}^{\uparrow\uparrow}ve^{iq_{+}a} 
\end{equation}
From Ref.~\onlinecite{AJP}, we have-
\begin{equation}
\vec s.\vec S=s^{Z}S^{Z}+\frac{1}{2}(s^{-}S^{+}+s^{+}S^{-}) 
\end{equation}
Here $s^{\pm} = s_{x}\pm is_{y}$ and $S^{\pm} = S_{x}\pm iS_{y}$ are the raising and lowering spin operators.\\
For spin up electron component
\begin{equation}
 \vec s.\vec S\begin{bmatrix}
               1\\
               0\\
               0\\
               0
              \end{bmatrix}\phi_{m'}^{S}=mm'\begin{bmatrix}
              1\\
              0\\
              0\\
              0
             \end{bmatrix}\phi_{m'}^{S}+\frac{1}{2}F_{1}F_{2}\begin{bmatrix}
             0\\
             1\\
             0\\
             0
             \end{bmatrix}\phi_{m'+1}^{S}\nonumber
\end{equation}
For spin down electron component
\begin{equation}
\vec s.\vec S\begin{bmatrix}
              0\\
              1\\
              0\\
              0\\
             \end{bmatrix}\phi_{m'+1}^{S}=(m-1)(m'+1)\begin{bmatrix}
             0\\
             1\\
             0\\
             0
             \end{bmatrix}\phi_{m'+1}^{S}+\frac{1}{2}F_{1}F_{2}\begin{bmatrix}
                                                                1\\
                                                                0\\
                                                                0\\
                                                                0
                                                               \end{bmatrix}\phi_{m'}^{S}\nonumber
\end{equation}
For spin up hole component
\begin{equation}
\vec s.\vec S\begin{bmatrix}
              0\\
              0\\
              1\\
              0
             \end{bmatrix}\phi_{m'+1}^{S}=(m-1)(m'+1)\begin{bmatrix}
             0\\
             0\\
             1\\
             0
             \end{bmatrix}\phi_{m'+1}^{S}+\frac{1}{2}F_{1}F_{2}\begin{bmatrix}
             0\\
             0\\
             0\\
             1
             \end{bmatrix}\phi_{m'}^{S}\nonumber
\end{equation}
For spin down hole component
\begin{equation}
\vec s.\vec S\begin{bmatrix}
              0\\
              0\\
              0\\
              1
             \end{bmatrix}\phi_{m'}^{S}=mm'\begin{bmatrix}
             0\\
             0\\
             0\\
             1
            \end{bmatrix}\phi_{m'}^{S}+\frac{1}{2}F_{1}F_{2}\begin{bmatrix}
            0\\
            0\\
            1\\
            0
            \end{bmatrix}\phi_{m'+1}^{S}\nonumber
\end{equation}
where $m$ and $m'$ are the electron spin and impurity spin respectively. $F_{1}=\sqrt{(s+m)(s-m+1)}$ and $F_{2}=\sqrt{(S-m')(S+m'+1)}$. Now
from boundary condition $(15)$ after some algebraic manipulation we get
\begin{equation}
ik_{e}[t_{ee}^{'\uparrow\uparrow}-b_{ee}^{\uparrow\uparrow}e^{ik_{e}a}-1+r_{ee}^{\uparrow\uparrow}]=-(m^{\star}J_{0}/\hbar^2)[2mm'(1+r_{ee}^{\uparrow\uparrow})+F_{1}F_{2}r_{ee}^{\uparrow\downarrow}] 
\end{equation}
\begin{equation}
ik_{e}[t_{ee}^{'\uparrow\downarrow}-b_{ee}^{\uparrow\downarrow}e^{ik_{e}a}+r_{ee}^{\uparrow\downarrow}]=-(m^{\star}J_{0}/\hbar^2)[2(m-1)(m'+1)r_{ee}^{\uparrow\downarrow}+F_{1}F_{2}(1+r_{ee}^{\uparrow\uparrow})] 
\end{equation}
\begin{equation}
ik_{h}[c_{eh}^{\uparrow\uparrow}e^{-ik_{h}a}-a_{eh}^{\uparrow\uparrow}-r_{eh}^{\uparrow\uparrow}]=-(m^{\star}J_{0}/\hbar^2)[2(m-1)(m'+1)r_{eh}^{\uparrow\uparrow}+F_{1}F_{2}r_{eh}^{\uparrow\downarrow}]
\end{equation}
\begin{equation}
ik_{h}[c_{eh}^{\uparrow\downarrow}e^{-ik_{h}a}-a_{eh}^{\uparrow\downarrow}-r_{eh}^{\uparrow\downarrow}]=-(m^{\star}J_{0}/\hbar^2)[2mm'r_{eh}^{\uparrow\downarrow}+F_{1}F_{2}r_{eh}^{\uparrow\uparrow}]
\end{equation}
From boundary condition $(17)$ after some algebraic calculation we get
\begin{equation}
iq_{+}t_{ee}^{\uparrow\uparrow}ue^{iq_{+}a}-iq_{-}t_{eh}^{\uparrow\downarrow}ve^{-iq_{-}a}-ik_{e}t_{ee}^{'\uparrow\uparrow}e^{ik_{e}a}+ik_{e}b_{ee}^{\uparrow\uparrow}=(2m^{\star}V/\hbar^2)[t_{ee}^{'\uparrow\uparrow}e^{ik_{e}a}+b_{ee}^{\uparrow\uparrow}] 
\end{equation}
\begin{equation}
iq_{+}t_{ee}^{\uparrow\downarrow}ue^{iq_{+}a}+iq_{-}t_{eh}^{\uparrow\uparrow}ve^{-iq_{-}a}-ik_{e}t_{ee}^{'\uparrow\downarrow}e^{ik_{e}a}+ik_{e}b_{ee}^{\uparrow\downarrow}=(2m^{\star}V/\hbar^2)[t_{ee}^{'\uparrow\downarrow}e^{ik_{e}a}+b_{ee}^{\uparrow\downarrow}] 
\end{equation}
\begin{equation}
-iq_{+}t_{ee}^{\uparrow\downarrow}ve^{iq_{+}a}-iq_{-}t_{eh}^{\uparrow\uparrow}ue^{-iq_{-}a}-ik_{h}c_{eh}^{\uparrow\uparrow}+ik_{h}a_{eh}^{\uparrow\uparrow}e^{-ik_{h}a}=(2m^{\star}V/\hbar^2)[c_{eh}^{\uparrow\uparrow}+a_{eh}^{\uparrow\uparrow}e^{-ik_{h}a}] 
\end{equation}
\begin{equation}
iq_{+}t_{ee}^{\uparrow\uparrow}ve^{iq_{+}a}-iq_{-}t_{eh}^{\uparrow\downarrow}ue^{-iq_{-}a}-ik_{h}c_{eh}^{\uparrow\downarrow}+ik_{h}a_{eh}^{\uparrow\downarrow}e^{-ik_{h}a}=(2m^{\star}V/\hbar^2)[c_{eh}^{\uparrow\downarrow}+a_{eh}^{\uparrow\downarrow}e^{-ik_{h}a}] 
\end{equation}
In Andreev approximation we take $k_{e}=k_{h}=q_{+}=q_{-}=k_{F}$. So finally we get
\begin{eqnarray}
r_{ee}^{\uparrow\uparrow}-t_{ee}^{'\uparrow\uparrow}-b_{ee}^{\uparrow\uparrow}e^{ik_{F}a}=-1,\nonumber\\
r_{ee}^{\uparrow\downarrow}-t_{ee}^{'\uparrow\downarrow}-b_{ee}^{\uparrow\downarrow}e^{ik_{F}a}=0,\nonumber\\ 
r_{eh}^{\uparrow\uparrow}-c_{eh}^{\uparrow\uparrow}e^{-ik_{F}a}-a_{eh}^{\uparrow\uparrow}=0,\nonumber\\ 
r_{eh}^{\uparrow\downarrow}-c_{eh}^{\uparrow\downarrow}e^{-ik_{F}a}-a_{eh}^{\uparrow\downarrow}=0,\nonumber\\ 
t_{ee}^{'\uparrow\uparrow}e^{ik_{F}a}+b_{ee}^{\uparrow\uparrow}-t_{ee}^{\uparrow\uparrow}ue^{ik_{F}a}-t_{eh}^{\uparrow\downarrow}ve^{-ik_{F}a}=0,\nonumber\\ 
t_{ee}^{'\uparrow\downarrow}e^{ik_{F}a}+b_{ee}^{\uparrow\downarrow}+t_{eh}^{\uparrow\uparrow}ve^{-ik_{F}a}-t_{ee}^{\uparrow\downarrow}ue^{ik_{F}a}=0,\nonumber\\ 
c_{eh}^{\uparrow\uparrow}+a_{eh}^{\uparrow\uparrow}e^{-ik_{F}a}-t_{eh}^{\uparrow\uparrow}ue^{-ik_{F}a}+   t_{ee}^{\uparrow\downarrow}ve^{ik_{F}a}=0,\nonumber\\ 
c_{eh}^{\uparrow\downarrow}+a_{eh}^{\uparrow\downarrow}e^{-ik_{F}a}-t_{ee}^{\uparrow\uparrow}ve^{ik_{F}a}-t_{eh}^{\uparrow\downarrow}ue^{-ik_{F}a}=0,\nonumber\\ 
(1-i2Jmm')r_{ee}^{\uparrow\uparrow}-iJF_{1}F_{2}r_{ee}^{\uparrow\downarrow}+t_{ee}^{'\uparrow\uparrow}-b_{ee}^{\uparrow\uparrow}e^{ik_{F}a}=(1+i2Jmm'),\nonumber\\ 
iJF_{1}F_{2}r_{ee}^{\uparrow\uparrow}+(i2J(m-1)(m'+1)-1)r_{ee}^{\uparrow\downarrow}-t_{ee}^{'\uparrow\downarrow} +b_{ee}^{\uparrow\downarrow}e^{ik_{F}a}=-iJF_{1}F_{2},\nonumber\\
(1+i2J(m-1)(m'+1))r_{eh}^{\uparrow\uparrow}+iJF_{1}F_{2}r_{eh}^{\uparrow\downarrow}-e^{-ik_{F}a}c_{eh}^{\uparrow\uparrow}+a_{eh}^{\uparrow\uparrow}=0,\nonumber\\ 
iJF_{1}F_{2}r_{eh}^{\uparrow\uparrow}+(1+i2Jmm')r_{eh}^{\uparrow\downarrow}-c_{eh}^{\uparrow\downarrow}e^{-ik_{F}a}+a_{eh}^{\uparrow\downarrow}=0,\nonumber\\ 
(i2Z-1)e^{ik_{F}a}t_{ee}^{'\uparrow\uparrow}+(1+i2Z)b_{ee}^{\uparrow\uparrow}+t_{ee}^{\uparrow\uparrow}ue^{ik_{F}a}-t_{eh}^{\uparrow\downarrow}ve^{-ik_{F}a}=0,\nonumber\\
(i2Z-1)e^{ik_{F}a}t_{ee}^{'\uparrow\downarrow}+(1+i2Z)b_{ee}^{\uparrow\downarrow}+t_{ee}^{\uparrow\downarrow}ue^{ik_{F}a}+
t_{eh}^{\uparrow\uparrow}ve^{-ik_{F}a}=0,\nonumber\\ 
(i2Z-1)c_{eh}^{\uparrow\uparrow}+(i2Z+1)e^{-ik_{F}a}a_{eh}^{\uparrow\uparrow}-t_{eh}^{\uparrow\uparrow}ue^{-ik_{F}a}-t_{ee}^{\uparrow\downarrow}ve^{ik_{F}a}=0,\nonumber\\ 
(i2Z-1)c_{eh}^{\uparrow\downarrow}+(i2Z+1)e^{-ik_{F}a}a_{eh}^{\uparrow\downarrow}-t_{eh}^{\uparrow\downarrow}ue^{-ik_{F}a}+t_{ee}^{\uparrow\uparrow}ve^{ik_{F}a}=0.
\end{eqnarray}
where $J = \frac{m^{\star}J_{0}}{\hbar^2k_{F}}$ and $Z = \frac{m^{\star}V}{\hbar^2k_{F}}$. We solve the above 16 equations in Eq.~(35) to calculate the different reflection and transmission probabilities $R_{ee}^{\uparrow\uparrow}= |r_{ee}^{\uparrow\uparrow}|^{2},R_{ee}^{\uparrow\downarrow}=|r_{ee}^{\uparrow\downarrow}|^{2},R_{eh}^{\uparrow\uparrow}=|r_{eh}^{\uparrow\uparrow}|^{2},R_{eh}^{\uparrow\downarrow}=|r_{eh}^{\uparrow\downarrow}|^{2}$, {$T_{ee}^{\uparrow\uparrow}=(u^2-v^2)|t_{ee}^{\uparrow\uparrow}|^{2},T_{ee}^{\uparrow\downarrow}=(u^2-v^2)|t_{ee}^{\uparrow\downarrow}|^{2},T_{eh}^{\uparrow\uparrow}=(u^2-v^2)|t_{eh}^{\uparrow\uparrow}|^{2},T_{eh}^{\uparrow\downarrow}=(u^2-v^2)|t_{eh}^{\uparrow\downarrow}|^{2}$}.
\newpage
\subsection{Table I}
\begin{table}[h]
\caption{$G_{c}$ and $G_{s}$ values for different $S$ and $m'$ for $E=0, J=1, Z=0$}
\label{aggiungi}\centering%
\begin{tabular}{ c  c  c  c  c  c }
\toprule%
$S$\hspace{10ex} & $m'$\hspace{10ex} & $F_{2}$\hspace{10ex} & $F_{4}$\hspace{10ex} & $G_{c}$\hspace{10ex} & $G_{s}$\\\hline \vspace{5px}
$\frac{1}{2}$\hspace{10ex} &$-\frac{1}{2}$\hspace{10ex} &$1$\hspace{10ex} &$0$\hspace{10ex} &$3.01845$\hspace{10ex} &$1.4382$\\\vspace{5px}
              &$\frac{1}{2}$\hspace{9ex}  &$0$\hspace{10ex} &$1$\hspace{10ex} &$3.01845$\hspace{10ex} &$-1.4382$\\\hline\vspace{5px}
$\frac{3}{2}$\hspace{10ex} &$-\frac{3}{2}$\hspace{10ex} &$\sqrt{3}$\hspace{10ex} &$0$\hspace{10ex} &$0.812352$\hspace{10ex} &$1.04065$\\\vspace{5px}
              &$-\frac{1}{2}$\hspace{10ex}              &$2$\hspace{10ex}        &$\sqrt{3}$\hspace{10ex} &$0.804158$\hspace{10ex} &$0.638038$\\\vspace{5px}
              &$\frac{1}{2}$\hspace{9ex}  &$\sqrt{3}$\hspace{10ex}  &$2$\hspace{10ex}       &$0.804158$\hspace{10ex} &$-0.638038$\\\vspace{5px}
              &$\frac{3}{2}$\hspace{9ex}  &$0$\hspace{10ex}        &$\sqrt{3}$\hspace{10ex} &$0.812352$\hspace{10ex} &$-1.04065$\\\hline\vspace{5px}
$\frac{5}{2}$\hspace{10ex} &$-\frac{5}{2}$\hspace{10ex} &$\sqrt{5}$\hspace{10ex} &$0$\hspace{10ex}        &$0.202059$\hspace{10ex} &$0.409994$\\\vspace{5px}
              &$-\frac{3}{2}$\hspace{10ex} &$2\sqrt{2}$\hspace{10ex} &$\sqrt{5}$\hspace{10ex} &$0.198990$\hspace{10ex} & $0.337998$\\\vspace{5px}
              &$-\frac{1}{2}$\hspace{10ex} &$3$\hspace{10ex}        &$2\sqrt{2}$\hspace{10ex} &$0.183388$\hspace{10ex} &$0.134767$\\\vspace{5px}
              &$\frac{1}{2}$\hspace{9ex}  &$2\sqrt{2}$\hspace{10ex} &$3$\hspace{10ex}        &$0.183388$\hspace{10ex} &$-0.134767$\\\vspace{5px}
              &$\frac{3}{2}$\hspace{9ex}  &$\sqrt{5}$\hspace{10ex} &$2\sqrt{2}$\hspace{10ex} &$0.198990$\hspace{10ex} &$-0.337998$\\\vspace{5px}
              &$\frac{5}{2}$\hspace{9ex}  &$0$\hspace{10ex}        &$\sqrt{5}$\hspace{10ex}  &$0.202059$\hspace{10ex} &$-0.409994$\\\hline \vspace{5px}
$\frac{7}{2}$\hspace{10ex} &$-\frac{7}{2}$\hspace{10ex} &$\sqrt{7}$\hspace{10ex} &$0$\hspace{10ex} &$0.0663543$\hspace{10ex} &$0.193730$\\\vspace{5px}
              &$-\frac{5}{2}$\hspace{10ex} &$2\sqrt{3}$\hspace{10ex} &$\sqrt{7}$\hspace{10ex} &$0.064999$\hspace{10ex} &$0.163065$\\\vspace{5px}
              &$-\frac{3}{2}$\hspace{10ex} &$\sqrt{15}$\hspace{10ex} &$2\sqrt{3}$\hspace{10ex} &$0.0622832$\hspace{10ex} &$0.110018$\\\vspace{5px}
              &$-\frac{1}{2}$\hspace{10ex} &$4$\hspace{10ex}         &$\sqrt{15}$\hspace{10ex} &$0.0601307$\hspace{10ex} &$0.0389914$\\\vspace{5px}
              &$\frac{1}{2}$\hspace{9ex}  &$\sqrt{15}$\hspace{10ex} &$4$\hspace{10ex}         &$0.0601307$\hspace{10ex} &$-0.0389914$\\\vspace{5px}
              &$\frac{3}{2}$\hspace{9ex}  &$2\sqrt{3}$\hspace{10ex} &$\sqrt{15}$\hspace{10ex} &$0.0622832$\hspace{10ex} &$-0.110018$\\\vspace{5px}
              &$\frac{5}{2}$\hspace{9ex}  &$\sqrt{7}$\hspace{10ex}  &$2\sqrt{3}$\hspace{10ex}  &$0.064999$\hspace{10ex} &$-0.163065$\\\vspace{5px}
              &$\frac{7}{2}$\hspace{9ex}  &$0$\hspace{10ex}         &$\sqrt{7}$\hspace{10ex}   &$0.0663543$\hspace{10ex} &$-0.193730$\\\hline \vspace{5px}
$\frac{9}{2}$\hspace{10ex} &$-\frac{9}{2}$\hspace{10ex} &$3$\hspace{10ex} &$0$\hspace{10ex}    &$0.0271740$\hspace{10ex} &$0.105640$\\\vspace{5px}
                           &$-\frac{7}{2}$\hspace{10ex} &$4$\hspace{10ex} &$3$\hspace{10ex}    &$0.0266748$\hspace{10ex} &$0.0902101$\\\vspace{5px}
                           &$-\frac{5}{2}$\hspace{10ex} &$\sqrt{21}$\hspace{10ex} &$4$\hspace{10ex} &$0.0260063$\hspace{10ex} &$0.0693175$\\\vspace{5px}
                           &$-\frac{3}{2}$\hspace{10ex} &$2\sqrt{6}$\hspace{10ex} &$\sqrt{21}$\hspace{10ex} &$0.0253821$\hspace{10ex} &$0.0437317$\\\vspace{5px}
                           &$-\frac{1}{2}$\hspace{10ex} &$5$\hspace{10ex} &$2\sqrt{6}$\hspace{10ex} &$0.025007$\hspace{10ex} &$0.0149549$\\\vspace{5px}
                           &$\frac{1}{2}$\hspace{9ex} &$2\sqrt{6}$\hspace{10ex} &$5$\hspace{10ex} &$0.025007$\hspace{10ex} &$-0.0149549$\\\vspace{5px}
                           &$\frac{3}{2}$\hspace{9ex} &$\sqrt{21}$\hspace{10ex} &$2\sqrt{6}$\hspace{10ex} &$0.0253821$\hspace{10ex} &$-0.0437317$\\\vspace{5px}
                           &$\frac{5}{2}$\hspace{9ex} &$4$\hspace{10ex} &$\sqrt{21}$\hspace{10ex} &$0.0260063$\hspace{10ex} &$-0.0693175$\\\vspace{5px}
                           &$\frac{7}{2}$\hspace{9ex} &$3$\hspace{10ex} &$4$\hspace{10ex} &$0.0266748$\hspace{10ex} &$-0.0902101$\\\vspace{5px}
                           &$\frac{9}{2}$\hspace{9ex} &$0$\hspace{10ex} &$3$\hspace{10ex} &$0.0271740$\hspace{10ex} &$-0.105640$\\\hline
\end{tabular} 
\end{table}
\newpage
\subsection{Table II}
\begin{table}[h]
\caption{$G_{c}$ and $G_{s}$ values for different $S$ and $m'$ for $E=\Delta, J=1, Z=0$}
\label{aggiungi}\centering%
\begin{tabular}{ c  c  c  c  c  c }
\toprule%
$S$\hspace{10ex} & $m'$\hspace{10ex} & $F_{2}$\hspace{10ex} & $F_{4}$\hspace{10ex} & $G_{c}$\hspace{10ex} & $G_{s}$\\\hline \vspace{5px}
$\frac{1}{2}$\hspace{10ex} &$-\frac{1}{2}$\hspace{10ex} &$1$\hspace{10ex} &$0$\hspace{10ex} &$3$\hspace{10ex} &$1$\\\vspace{5px}
              &$\frac{1}{2}$\hspace{9ex}  &$0$\hspace{10ex} &$1$\hspace{10ex} &$3$\hspace{10ex} &$-1$\\\hline\vspace{5px}
$\frac{3}{2}$\hspace{10ex} &$-\frac{3}{2}$\hspace{10ex} &$\sqrt{3}$\hspace{10ex} &$0$\hspace{10ex} &$2.5$\hspace{10ex} &$0.75$\\\vspace{5px}
              &$-\frac{1}{2}$\hspace{10ex}              &$2$\hspace{10ex}        &$\sqrt{3}$\hspace{10ex} &$0.9$\hspace{10ex} &$-0.11$\\\vspace{5px}
              &$\frac{1}{2}$\hspace{9ex}  &$\sqrt{3}$\hspace{10ex}  &$2$\hspace{10ex}       &$0.9$\hspace{10ex} &$0.11$\\\vspace{5px}
              &$\frac{3}{2}$\hspace{9ex}  &$0$\hspace{10ex}        &$\sqrt{3}$\hspace{10ex} &$2.5$\hspace{10ex} &$-0.75$\\\hline\vspace{5px}
$\frac{5}{2}$\hspace{10ex} &$-\frac{5}{2}$\hspace{10ex} &$\sqrt{5}$\hspace{10ex} &$0$\hspace{10ex}        &$2.33333$\hspace{10ex} &$0.555556$\\\vspace{5px}
              &$-\frac{3}{2}$\hspace{10ex} &$2\sqrt{2}$\hspace{10ex} &$\sqrt{5}$\hspace{10ex} &$0.555556$\hspace{10ex} & $-0.160494$\\\vspace{5px}
              &$-\frac{1}{2}$\hspace{10ex} &$3$\hspace{10ex}        &$2\sqrt{2}$\hspace{10ex} &$0.422222$\hspace{10ex} &$-0.0350617$\\\vspace{5px}
              &$\frac{1}{2}$\hspace{9ex}  &$2\sqrt{2}$\hspace{10ex} &$3$\hspace{10ex}        &$0.422222$\hspace{10ex} &$0.0350617$\\\vspace{5px}
              &$\frac{3}{2}$\hspace{9ex}  &$\sqrt{5}$\hspace{10ex} &$2\sqrt{2}$\hspace{10ex} &$0.555556$\hspace{10ex} &$0.160494$\\\vspace{5px}
              &$\frac{5}{2}$\hspace{9ex}  &$0$\hspace{10ex}        &$\sqrt{5}$\hspace{10ex}  &$2.33333$\hspace{10ex} &$-0.555556$\\\hline \vspace{5px}
$\frac{7}{2}$\hspace{10ex} &$-\frac{7}{2}$\hspace{10ex} &$\sqrt{7}$\hspace{10ex} &$0$\hspace{10ex} &$2.25$\hspace{10ex} &$0.4375$\\\vspace{5px}
              &$-\frac{5}{2}$\hspace{10ex} &$2\sqrt{3}$\hspace{10ex} &$\sqrt{7}$\hspace{10ex} &$0.403846$\hspace{10ex} &$-0.153476$\\\vspace{5px}
              &$-\frac{3}{2}$\hspace{10ex} &$\sqrt{15}$\hspace{10ex} &$2\sqrt{3}$\hspace{10ex} &$0.278846$\hspace{10ex} &$-0.0496487$\\\vspace{5px}
              &$-\frac{1}{2}$\hspace{10ex} &$4$\hspace{10ex}         &$\sqrt{15}$\hspace{10ex} &$0.242647$\hspace{10ex} &$-0.0129217$\\\vspace{5px}
              &$\frac{1}{2}$\hspace{9ex}  &$\sqrt{15}$\hspace{10ex} &$4$\hspace{10ex}         &$0.242647$\hspace{10ex} &$0.0129217$\\\vspace{5px}
              &$\frac{3}{2}$\hspace{9ex}  &$2\sqrt{3}$\hspace{10ex} &$\sqrt{15}$\hspace{10ex} &$0.278846$\hspace{10ex} &$0.0496487$\\\vspace{5px}
              &$\frac{5}{2}$\hspace{9ex}  &$\sqrt{7}$\hspace{10ex}  &$2\sqrt{3}$\hspace{10ex}  &$0.403846$\hspace{10ex} &$0.153476$\\\vspace{5px}
              &$\frac{7}{2}$\hspace{9ex}  &$0$\hspace{10ex}         &$\sqrt{7}$\hspace{10ex}   &$2.25$\hspace{10ex} &$-0.4375$\\\hline \vspace{5px}
$\frac{9}{2}$\hspace{10ex} &$-\frac{9}{2}$\hspace{10ex} &$3$\hspace{10ex} &$0$\hspace{10ex}    &$2.2$\hspace{10ex} &$0.36$\\\vspace{5px}
                           &$-\frac{7}{2}$\hspace{10ex} &$4$\hspace{10ex} &$3$\hspace{10ex}    &$0.317647$\hspace{10ex} &$-0.138547$\\\vspace{5px}
                           &$-\frac{5}{2}$\hspace{10ex} &$\sqrt{21}$\hspace{10ex} &$4$\hspace{10ex} &$0.208556$\hspace{10ex} &$-0.0478996$\\\vspace{5px}
                           &$-\frac{3}{2}$\hspace{10ex} &$2\sqrt{6}$\hspace{10ex} &$\sqrt{21}$\hspace{10ex} &$0.170909$\hspace{10ex} &$-0.0199537$\\\vspace{5px}
                           &$-\frac{1}{2}$\hspace{10ex} &$5$\hspace{10ex} &$2\sqrt{6}$\hspace{10ex} &$0.156923$\hspace{10ex} &$-0.00567101$\\\vspace{5px}
                           &$\frac{1}{2}$\hspace{9ex} &$2\sqrt{6}$\hspace{10ex} &$5$\hspace{10ex} &$0.156923$\hspace{10ex} &$0.00567101$\\\vspace{5px}
                           &$\frac{3}{2}$\hspace{9ex} &$\sqrt{21}$\hspace{10ex} &$2\sqrt{6}$\hspace{10ex} &$0.170909$\hspace{10ex} &$0.0199537$\\\vspace{5px}
                           &$\frac{5}{2}$\hspace{9ex} &$4$\hspace{10ex} &$\sqrt{21}$\hspace{10ex} &$0.208556$\hspace{10ex} &$0.0478996$\\\vspace{5px}
                           &$\frac{7}{2}$\hspace{9ex} &$3$\hspace{10ex} &$4$\hspace{10ex} &$0.317647$\hspace{10ex} &$0.138547$\\\vspace{5px}
                           &$\frac{9}{2}$\hspace{9ex} &$0$\hspace{10ex} &$3$\hspace{10ex} &$2.2$\hspace{10ex} &$-0.36$\\\hline
\end{tabular} 
\end{table}
\section{Acknowledgements }
This work was supported by the grant ``Non-local correlations in nanoscale systems: Role of decoherence, interactions, disorder and pairing symmetry'' from SCIENCE \& ENGINEERING RESEARCH BOARD, New Delhi, Government of India, Grant No.  EMR/20l5/001836, Principal Investigator: Dr. Colin Benjamin, National Institute of Science Education and Research, Bhubaneswar, India.

\end{document}